\documentclass[twocolumn]{aastex631}

\begin{document}

\title{VLT/MUSE Characterisation of Dimorphos Ejecta from the DART Impact \footnote{Accepted November 6, 2023}}

\author[0000-0002-8137-5132]{Brian P. Murphy}
\affiliation{University of Edinburgh, Institute for Astronomy, Royal Observatory Edinburgh EH9 3HJ, UK}
\email{brian.murphy@ed.ac.uk}
\correspondingauthor{Brian Murphy}

\author[0000-0002-9298-7484]{Cyrielle Opitom}
\affiliation{University of Edinburgh, Institute for Astronomy, Royal Observatory Edinburgh EH9 3HJ, UK}

\author[0000-0001-9328-2905]{Colin Snodgrass}
\affiliation{University of Edinburgh, Institute for Astronomy, Royal Observatory Edinburgh EH9 3HJ, UK}

\author[0000-0003-2781-6897]{Matthew M. Knight}
\affiliation{United States Naval Academy Annapolis, MD, USA}

\author[0000-0003-3841-9977]{Jian-Yang Li}
\affiliation{Planetary Science Institute Tuscon, Arizona, USA}

\author[0000-0001-8628-3176]{Nancy L. Chabot}
\affiliation{Johns Hopkins University Applied Physics Laboratory, Laurel, MD, USA}

\author[0000-0002-9939-9976]{Andrew S. Rivkin}
\affiliation{Johns Hopkins University Applied Physics Laboratory, Laurel, MD, USA}

\author[0000-0002-9153-9786]{Simon F. Green}
\affiliation{The Open University School of Physical Sciences, Milton Keynes MK7 6AA, UK}

\author[0009-0002-6978-7377]{Paloma Guetzoyan}
\affiliation{University of Edinburgh, Institute for Astronomy, Royal Observatory Edinburgh EH9 3HJ, UK}

\author[0000-0002-9925-0426]{Daniel Gardener}
\affiliation{University of Edinburgh, Institute for Astronomy, Royal Observatory Edinburgh EH9 3HJ, UK}

\author[0000-0002-0696-0411]{Julia de León}
\affiliation{Instituto de Astrofísica de Canarias (IAC) C/Vía Láctea s/n, E-38205 La Laguna, Spain}
\affiliation{University of La Laguna, Department of Astrophysics, Tenerife, Spain}

\begin{abstract}

We have observed the Didymos-Dimorphos binary system with the MUSE integral field unit spectrograph mounted at the Very Large Telescope (VLT) pre and post-DART impact, and captured the ensuing ejecta cone, debris cloud, and tails at sub-arcsecond resolutions. We targeted the Didymos system over 11 nights from 26 September to 25 October 2022, and utilized both narrow and wide-field observations with and without adaptive optics, respectively. We took advantage of the spectral-spatial coupled measurements and produced both white-light images and spectral maps of the dust reflectance. We identified and characterized numerous dust features, such as the ejecta cone, spirals, wings, clumps, and tails. We found that the base of the Sunward edge of the wings, from 03 to 19 October, consistent with maximum grain sizes on the order of 0.05-0.2 mm, and that the earliest detected clumps have the highest velocities on the order of 10 m s$^{-1}$. We also see that three clumps in narrow-field mode (8x8$\arcsec$) exhibit redder colors and slower speeds, around 0.09 m s$^{-1}$, than the surrounding ejecta, likely indicating that the clump is comprised of larger, slower grains. We measured the properties of the primary tail, and resolved and measured the properties of the secondary tail earlier than any other published study, with first retrieval on 03 October. Both tails exhibit similarities in curvature and relative flux, however, the secondary tail appears thinner, which may be caused by lower energy ejecta and possibly a low energy formation mechanism such as secondary impacts.

\end{abstract}


\section{Introduction} \label{sec:intro}

The Double Asteroid Redirection Test (DART) mission represents a significant step forward in humanity's ongoing efforts to safeguard Earth from potential asteroid impacts and allowed us the opportunity to further understand the perturbation mechanisms of active asteroids, through the application of the kinetic redirect technique and controlled impact of the DART spacecraft \citep{Chabot:2023, Rivkin:2021, Cheng:2020}. With the primary aim of testing this technique and overall planetary defense strategies, the 579 kg DART spacecraft\footnote{https://dart.jhuapl.edu/Mission/Impactor-Spacecraft.php} impacted the 171$\pm$11 m \citep{Scheirich:2022} diameter asteroid Dimorphos on 26 September at 23:14 UT and excavated an estimated 1.3-2.2 $\times$ 10$^7$ kg of material, or 0.3-0.5$\%$ $M_\mathrm{Dimorphos}$ \citep{Daly:2023, Graykowski:2023}. The 6.1 km s$^{-1}$ impact imparted massive amounts of energy into the body, which perturbed Dimorphos into an active asteroid, increased observed brightness to m$_v$ = 12.18$\pm$0.03 \citep{Graykowski:2023}, and shortened Dimorphos' orbit around the larger 786$\pm$50 m \citep{Scheirich:2022} asteroid Didymos by 33.0$\pm$1.0 minutes (3$\sigma$) \citep{Thomas:2023}.  

Characterization of the ensuing ejecta cloud was coordinated across 59 ground-based observatories and 4 space-based facilities (Hubble Space Telescope, JWST, Lucy, LICIACube), from just seconds post-impact to almost a year after \citep{Chabot:2023}. Within the first minutes post-impact, fast ejecta from the vaporized DART spacecraft and Dimorphos' surface expanded spherically from the impact site at velocities around 1 - 3.6 km s$^{-1}$ \citep{Graykowski:2023, Fitzsimmons:2023, Weaver:2023, Shestakova:2023}. In the next few hours post-impact, the fast ejecta cleared the vicinity of the binary system, which allowed for in-depth photometric, spectroscopic, and morphological characterization of the impact ejecta. Further studies by \cite{Rozek:2023,Kareta:2023,Lin:2023, Li:2023} found that the system faded with time as ejecta left the binary system, although, a slight pause in fading was observed between 01 to 03 October (discussed in Section \ref{subsubsec:secondary}). \cite{Opitom:2023} used preliminary MUSE data to probe the ejecta cloud for any evidence of subsurface volatile species that may have been vaporized or excavated by the impact, and found that the ejecta was likely volatile-free. Finally, \cite{Li:2023, Opitom:2023, Lin:2023} extensively characterized the morphological evolution of the ejecta from just after impact to up to a month post-impact, which revealed numerous dust features such as the wide eastern-facing ejecta cone, diffuse surrounding ejecta cloud, various clumps and linea, spirals, wings, and the formation of two dust tails.

Each of these ejecta structures were primarily influenced by gravitational interactions and solar radiation pressure (SRP) \citep{Burns:1979} in the hours to days after impact. Gravitational interactions with the binary-dominated ejecta evolution for large slow particles (v$_\mathrm{esc}\leq$v$_\mathrm{particle}\leq$2v$_\mathrm{esc}$) in the vicinity of the system within 2 days post-impact, as suggested in models from \cite{Moreno:2022}. However, smaller faster particles were dominated by SRP-induced evolution and the contributions from SRP are dependent on the cross-sectional area of the dust particles, which directly correlates to mass if the dust particle density is assumed constant. Therefore, SRP preferentially influenced particles based on mass and size, and more rapidly accelerated and altered the trajectories of smaller lighter particles while larger heavier particles required longer timescales for SRP effects to become prominent. Models by \cite{Lin:2023} suggested SRP was the key force in the formation of the primary dust tail, which formed just hours post-impact and expanded at 31 m s$^{-1}$ in the anti-Sunward direction on 27 September. Finally, work by \cite{Li:2023} further established that by the 29 September Hubble Space Telescope (HST) observations, the majority of ejecta evolution was no longer governed by gravitational regimes, but rather by SRP-induced evolution.

Within four hours post-impact, the MUSE instrument at the VLT targeted the binary system and observed the expanding ejecta cone in sub-arcsecond resolution. Subsequently, MUSE revisited the system over 10 additional nights and produced 3D (x,y,$\lambda$) datacubes of the evolving ejecta until 25 October, nearly a month after impact. These observations provide crucial information regarding the physical characteristics, dynamics, ejection speeds, and evolution of the binary system post-impact, and have allowed us to better contextualize the impact event and aftermath of the DART mission. This paper will build upon initial findings from \cite{Opitom:2023} by presenting novel analyses across the entire observational dataset, which allows us to further infer correlations between the morphological features and spectral properties of the DART-induced Dimorphos ejecta. We report our ejecta cone and substructure results in Section \ref{subsec:cone}, primary tail measurements in Section \ref{subsubsec:primary}, and secondary tail measurements in Section \ref{subsubsec:secondary}. We did not carry out detailed modeling of our results, as it is outside the scope of this paper.

\section{Observations and Data Processing} \label{sec:observations and data processing}

    \begin{table*}[ht!]
        \caption{\label{TableObs} Observing Conditons throughout the 2I VLT/MUSE DART Campaign}
        \centering
        \begin{tabular}{lccccccccc}
        \hline\hline
        Date & Time (UT) & Mode & N$_\mathrm{raw}$ & N$_\mathrm{stacked}$ & $\Delta$T (d) & $\alpha$ (deg) & $\Delta$ (au) & Scale (km/$\arcsec$) & Seeing ($\arcsec$) \\
        \hline
        2022-Sep-26 & 07:51-09:08 & NFM+WFM & 6+4 & 1+1 & -0.63 & 52.3 & 0.076 & 55.4 & 0.66 \\
        2022-Sep-27 & 02:55-08:02 & NFM+WFM & 8+15 & 2+3 & +0.25 & 53.6 & 0.075 & 54.7 & 0.70 \\
        2022-Sep-28 & 03:54-08:17 & NFM+WFM & 6+14 & 1+3 & +1.29 & 55.2 & 0.074 & 53.9 & 1.10 \\
        2022-Sep-29 & 06:10-09:07 & NFM+WFM & 5+5 & 1+1 & +2.33 & 56.7 & 0.073 & 53.2 & 0.98 \\
        2022-Oct-01 & 06:30-09:08 & NFM+WFM & 5+4 & 1+1& +4.46 & 59.7 & 0.072 & 52.2 & 1.23 \\
        2022-Oct-03 & 07:22-09:09 & WFM & 10 & 1 & +6.38 & 62.7 & 0.071 & 51.7 & 0.63 \\
        2022-Oct-04 & 06:19-07:52 & WFM & 8 & 1 & +7.33 & 64.0 & 0.071 & 51.6 & 0.62 \\
        2022-Oct-07 & 05:51-09:31 & NFM+WFM & 5+10 & 1+1 & +10.33 & 67.9 & 0.071 & 52.1 & 0.36 \\
        2022-Oct-14 & 05:42-08:50 & WFM & 24 & 1 & +17.33 & 74.2 & 0.078 & 56.6 & 0.57 \\
        2022-Oct-19 & 05:31-08:59 & WFM & 25 & 1 & +22.33 & 76.1 & 0.085 & 62.1 & 1.72 \\
        2022-Oct-25 & 05:52-08:56 & WFM & 24 & 1 & +28.33 & 76.0 & 0.096 & 70.3 & 0.35 \\
        \hline
        \end{tabular}
        \newline
        \newline
        \textbf{Notes.}{ N$_\mathrm{raw}$ represents the number of exposures obtained, N$_\mathrm{stacked}$ is the number of final stacked images, $\Delta$T the number of hours between impact and the observations mid-point, $\alpha$ the phase-angle (Sun-comet-Earth), $\Delta$ is the geocentric distance, scale is sky projected distance per arcsecond, and seeing is the average atmospheric seeing of the observation window. \citep{Opitom:2023} \label{tab:obs}}
    \end{table*}

We utilized the Multi-Unit Spectroscopic Explorer (MUSE) instrument, mounted on the Very Large Telescope (VLT) Unit Telescope 4 (UT4), which is uniquely suited to characterize the impact due to it being comprised of integral field unit (IFU) spectrographs, which collected three-dimensional observations, two spatial dimensions (x,y) and one spectral dimension ($\lambda$). MUSE accomplished this through implementing various fore-optics to split the total MUSE field of view (FoV) into 24 subfields, which are then individually relayed and spectrally-dispersed into 24 identical IFU spectrographs and 4k-resolution E2V Deep Depletion MIT/LL CCDs \citep{Bacon:2010}. Utilizing MUSE, we directly correlated spectral information to spatial features -- which allowed for a more precise analysis of conditions post-impact.

\begin{figure*}[p]
        \begin{center}
        \includegraphics[scale=0.45]{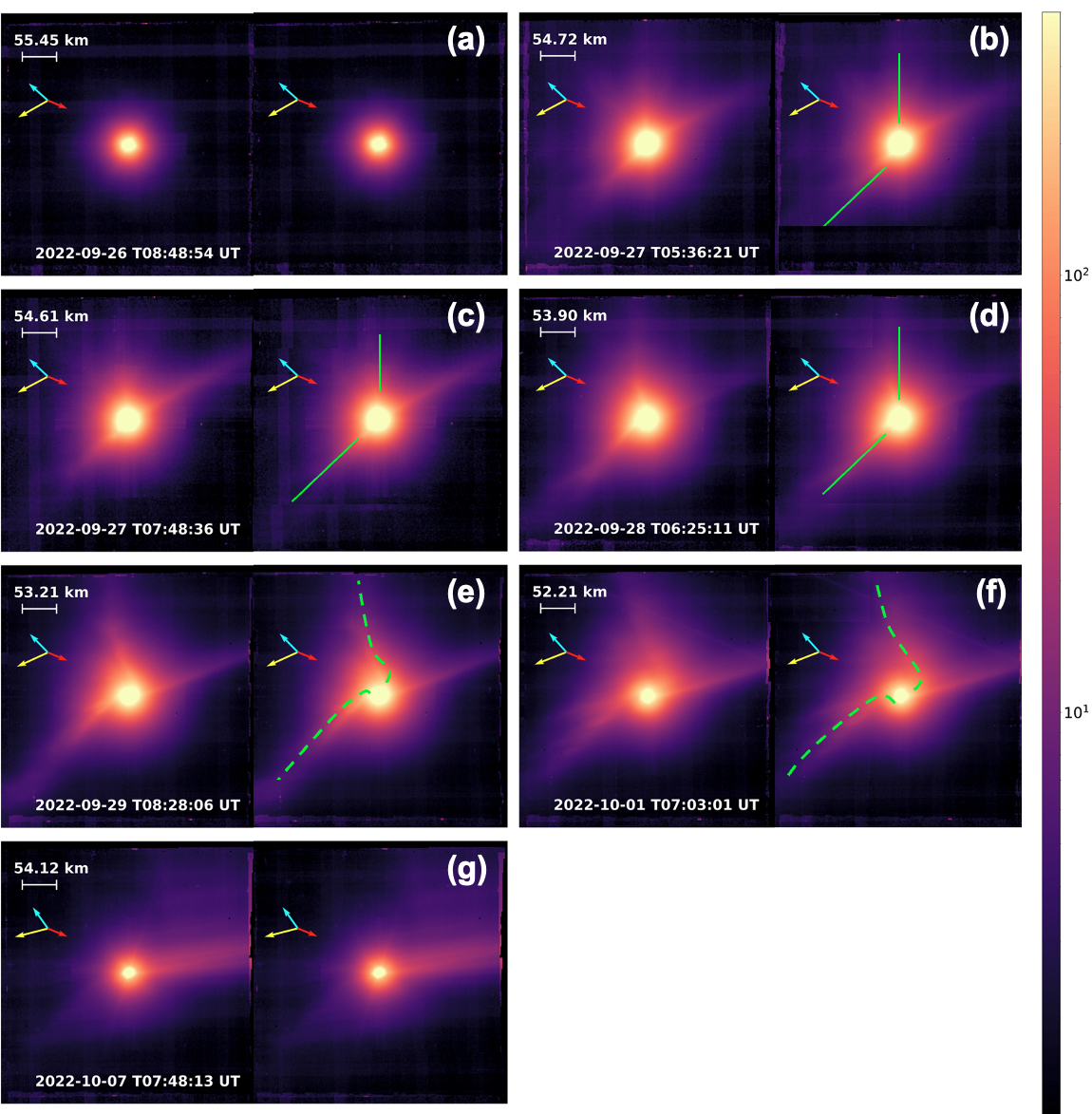}
        \caption{All white-light images from the NFM portion of our campaign, with AO online. Morphological evolution of the ejecta cone seen in panels (b) and (c), before gravitational interactions with the binary begin to curve the bases of the ejecta cone edges into spiral features by panels (d) and (e). By panel (f), the northern spiral has detached from the central system and propagates anti-Sunward until panel (g). Intensities are in units of 10$^{-20}$ erg/s/cm2. Convert to mks units using 1 erg = 10$^{-7}$J and 1 cm$^2$ = 0.0001 m$^2$. The yellow arrow indicates the projected angle toward the Sun, cyan arrow represents heliocentric orbital motion of the system, and red arrow is the DART spacecraft approach angle. Length of arrows are arbitrarily varied for clarity. \label{fig:NFM}}
        \end{center}
    \end{figure*}

\begin{figure*}[p]
        \begin{center}
        \includegraphics[scale=0.37]{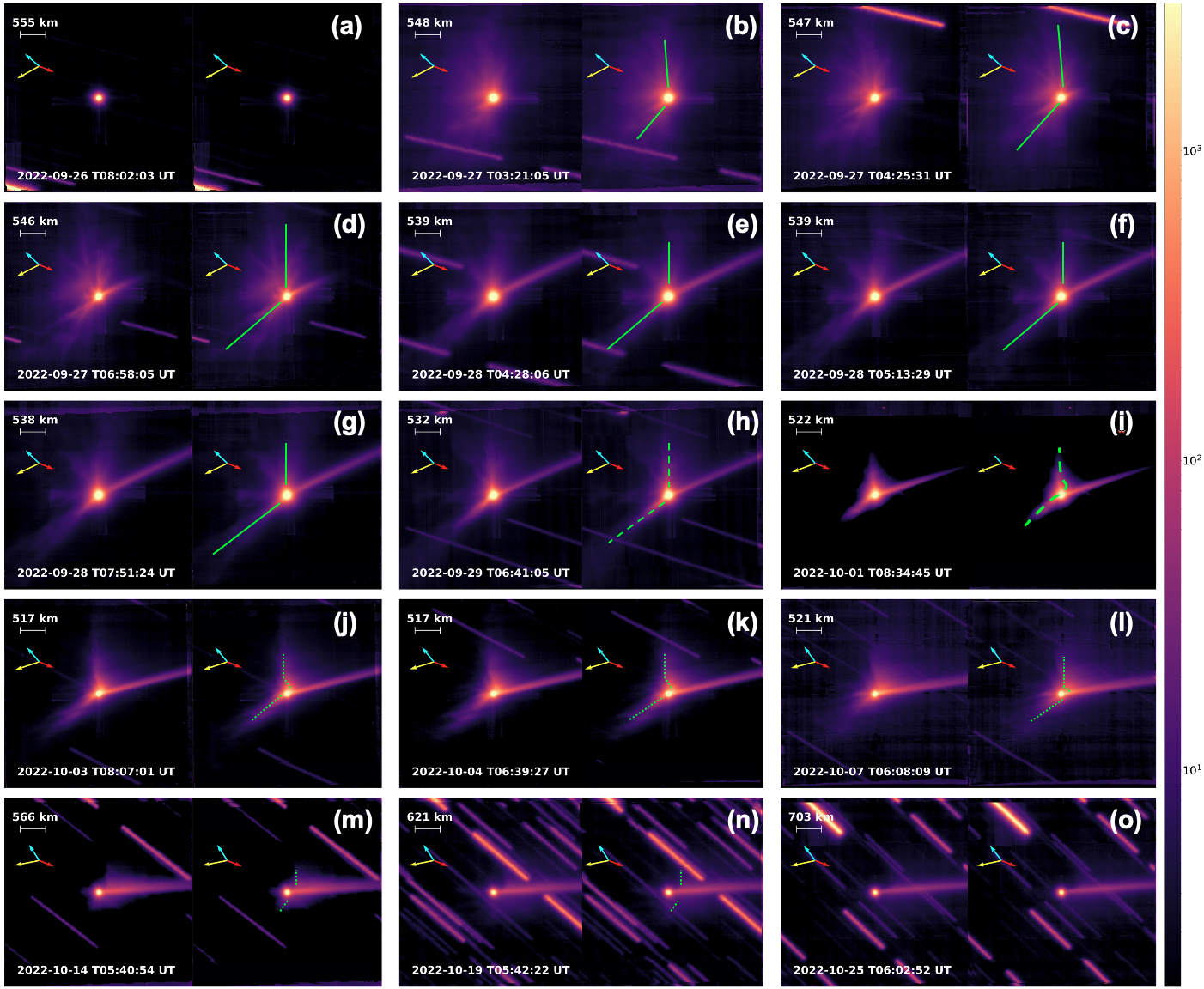}
        \caption{All white-light images from the main WFM portion of our campaign. Morphological evolution of the ejecta cone seen in panels (b) and (g), thereafter the spirals become increasingly visible until panel (k). By panel (l), the northern spiral has detached from the central system and propagates anti-Sunward until final detection in panel (n). Intensities are in units of 10$^{-20}$ erg/s/cm$^2$.\label{fig:WFM}}
        \end{center}
    \end{figure*}

\begin{figure*}[htb]
        \begin{center}
        \includegraphics[scale=0.45]{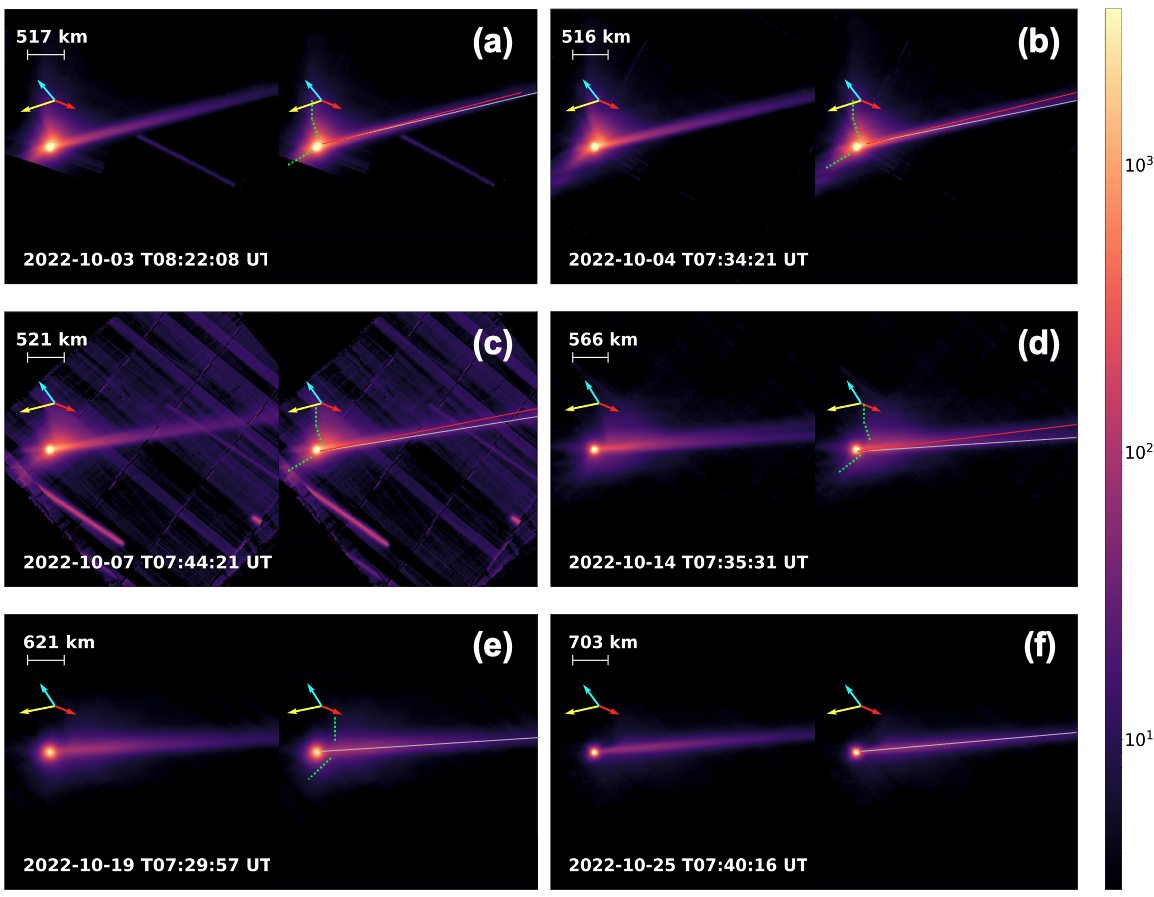}
        \caption{All offset white-light images from 03 to 25 October, in WFM. The anti-Sunward propagation of the wings can been seen more clearly in panels (c) through (e). The secondary tail becomes apparent at the outermost edge of the tail in panel (c), and is fully resolved along most of the tail in panel (d). The secondary tail is no longer observed in panels (e) and (f). Intensities are in units of 10$^{-20}$ erg/s/cm$^2$. \label{fig:WFM_offset}}
        \end{center}
    \end{figure*}

Ground-based MUSE observations were collected by our team as a part of ESO programmes 109.2361 and 110.23XL, from 26 September 2022 to 25 October 2022. We utilized MUSE in two observing modes -- Narrow Field Mode (NFM) with the Adaptive Optics Facility (AOF/GALACSI) and Wide Field Mode (WFM) without the AOF \citep{Arsenault:2008, Strobele:2012}. The sampling of MUSE is dependent on the observing mode, with NFM featuring an 8x8$\arcsec$ FoV and 0.025 $\arcsec$/spaxel sampling and WFM having a larger 60x60$\arcsec$ FoV and 0.2 $\arcsec$/spaxel sampling. The nominal spectral range covers the entire visible regime from 480 to 930-nm, and the spectral resolving power is dependent on wavelength and observing mode. The mean resolving powers for both NFM and WFM are around 3000, with a resolving power of 1740 at 480 nm and 3450 at 930 nm for NFM, and 1770 at 480 nm and 3590 at 930 nm for WFM \citep{Bacon:2010}. Since our NFM observations benefited from adaptive optics, we measured the quality of the turbulence-corrected observations by the width of the point spread function of the solar analogue star from each night, which estimated how well the AOF compensated for turbulence. The quality of our AO NFM observations are on the order of 100 mas (milliarcsecond) each night. The extent of our observations can be seen in Table \ref{tab:obs} \citep{Opitom:2023} and in Figures \ref{fig:NFM} through \ref{fig:WFM_offset}.

We employed exposure times that ranged from 300 to 600 seconds using non-sidereal tracking and captured faint details in the ejecta cloud without saturating the detector, such that shorter exposures coincided with brighter observations within days post-impact and longer observations were at later dates when the system was dimmer. We centered the observations on Didymos to capture the local environment, and also took offset and rotated observations from 03 October 2022 onward. These offset images centered Didymos in the South East quadrant of the detector, which better captured more of the tails and enhanced our spatial coverage. Our observations were made over a series of 11 nights and are comprised of 34 NFM and 144 WFM Didymos exposures. We designated 32 NFM and 139 WFM exposures for further analysis and rejected any exposures with evidence of tracking loss or excessive star-trail contamination. Of the 139 WFM exposures, 57 are offset. We ensured comprehensive coverage and mitigated instrumental effects between the 24 IFU detectors, by intentionally dithering and rotating all observations by $90^{\circ}$ from North to East between exposures. As part of the campaign, supplementary background sky observations (positioned 5$\arcsec$ to 10$\arcsec$ away from the system to limit source contamination), standard star, and solar analogue observations were also conducted for each observing mode. Observations were conducted in Object-Sky-Object-Object-Sky-Object pattern to ensure all observations of the system were temporally adjacent to a relevant background sky exposure. We observed solar analogue stars HD 28099 and HD 11532.

\subsection{Data Reduction} \label{sec:data reduction}

All exposures (Didymos, sky, solar analogue) were reduced using the ESO MUSE Pipeline \citep{Weilbacher:2020}. The pipeline utilized standard stars obtained by the observatory on the same night as science observations for flux calibration and telluric line correction, and performed wavelength calibration, computation of the Line Spread Function (LSF), instrument geometry and illumination corrections for the entire dataset. Background subtraction was treated independently for NFM and WFM data. NFM data were reduced using our dedicated sky observations due to the extended nature of the ejecta cloud that filled the entire FoV. The dedicated sky observations were processed by the pipeline assuming standard pixel fractions of 0.75 and 0.25 for clear sky and target, respectively, which means that 75\% of the image was assumed to have no source contamination and the remaining 25\% was assumed contaminated by signal from the source. Although these are dedicated sky observations and should have no source contamination, we retain redundancy by assuming 25\% source contamination. WFM data featured a larger FoV that allowed for the observations to contain sky background with no contamination from the source, so we tested modeled sky observations from the target frame against dedicated observations to deduce which would be sufficient for background subtraction. Lower sky residuals were achieved using sky backgrounds from the science frame, as shown by \cite{Opitom:2023}, so we proceeded with that background subtraction for all WFM data. The main products of the MUSE pipeline were flux-calibrated 3D datacubes and 2D white-light FoV images created from the full wavelength range of the datacubes.

\subsection{Image Processing} \label{sec:image proc}

This section describes the processing applied to the white light images produced by the MUSE pipeline. With the images rotated, reduced, and the data quality ensured through each of the reduction steps, we used two different stacking techniques to coadd all of the exposures. Using the MUSE Python Data Analysis Facility (MPDAF) from \cite{Piqueras:2017}, we coadded most of the white-light images over roughly 1h periods (20-75 minutes) for observations from 27 and 28 September, and over the entire night for all other dates. To ensure proper centering and reduce the effects of non-sidereal motion, we calculated the centroid of each frame using 2D Gaussian fits defined by 21x21 spaxel box around the primary source. We used MPDAF to implement dithering-shift corrections and coadded the exposures about the calculated centroids. White-light images from 14 through 25 October were treated with median coaddition in IRAF, which mitigated the increased star-trail contamination caused by Didymos' transit across the galactic plane. We did not use median stacking on other dates due to an insufficient amount of exposures, which were typically 2-6 raw exposures per coaddition and 15-20 per median addition. After this process, we were left with a total of 22 coadded science images across both MUSE modes.

We utilized the Comet Coma Image Enhancement Facility (CCIEF) \citep{Samarasinha:2013}, provided by the Planetary Science Institute\footnote{https://www.psi.edu/research/cometimen}, to elucidate the finer ejecta substructure in our science images by minimizing the effect of the surrounding diffuse debris cloud. We chose azimuthal average subtraction to remove the diffuse contributions of the debris cloud via radially calculating and subtracting the average background contribution in a series of concentric annuli. This method of averaging is less affected by outliers, such as bad pixels or remnant star trails, so that precise position angle measurements may be taken. Additionally, this enhancement method is considered benign and does not introduce artifacts, as demonstrated by \cite{Samarasinha:2014} and \cite{Schleicher:2004}. These images will be referred to as enhanced images for the remainder of this paper, and are featured in Figures \ref{fig:WFM_clump} and \ref{fig:NFM_clump}. The software converted the science and enhanced images into polar coordinates, such that rows and columns represent radial distance from Didymos and azimuthal angle around the asteroid system, from North to East, respectively. We produced 22 enhanced and 22 unenhanced polar projected images and Cartesian images. This process was implemented for all NFM and WFM science images. The subsequent morphological analysis in Section \ref{sec:method and results} utilized the original science images, enhanced polar and Cartesian images, and the unenhanced polar and Cartesian images.

\subsection{Datacube Processing} \label{sec:datacube proc}

The spectral information contained in the datacubes was used to study the spectral slope of the dust, similarly to what is presented in \cite{Opitom:2023}. In this study, we focus on the dust spectral slope in the NFM data, as WFM data were already presented in \cite{Opitom:2023}. For all NFM observations, we computed the relative reflectance spectrum normalised at 500~nm  for each spaxel of each datacube individually. We divided the spectrum of the object by a solar spectrum, computed the slope of the reflectance between 500 and 750~nm (this was shown to correspond to the peak of the reflectance spectrum of the dust ejecta by \cite{Opitom:2023}), and recorded the value of the slope for each spaxel, producing a spatial map of the dust reflectance slope. Spectral reflectance maps obtained within 1h30 of each other were averaged to produce final reflectance maps.

\section{Methodology and Results} \label{sec:method and results}

 We present the morphological and spectroscopic measurements taken from coadded datacubes and white-light images, and the methods of sampling for both NFM and WFM, in the following sections.

\subsection{Ejecta Cone Structure and Evolution} \label{subsec:cone}

We first detected the expanding ejecta cone on 27 September 2022 at 02:55 UT, or 0.15 days after impact (T+0.15D), as seen in Figure \ref{fig:WFM}. The sky projected ejecta cone faced east-north-east, with well-defined northern and southern ejecta cone edges. We report similar findings regarding ejecta cone structures, evolution, and morphology as \cite{Li:2023} and refer to Figures 2 and 3 in their work for an in-depth characterization of the ejecta structures and broader evolutionary patterns.

\subsubsection{Ejecta Cone Edges, Spirals, and Wings} \label{subsubsec:spirals}

From T+0.15D to T+1.29D after impact, the ejecta cone was defined by two collimated structures referred to as the northern edge and southern edge, seen as solid green lines in panels (b) through (d) in Figure \ref{fig:NFM} and (b) through (g) in Figure \ref{fig:WFM}. By 29 September, the effects of gravitational interactions with the binary manifested as clockwise curvature at the base of the edges, hereafter referred to as the northern and southern spirals \citep{Li:2023}, illustrated by dashed green lines in panels (e) to (f) in Figure \ref{fig:NFM} and (h) to (i) in Figure \ref{fig:WFM}. This clockwise curvature is a product of extended processing by the binary gravitational field in the vicinity of the system. Only the base of the spirals were significantly curved because they were comprised of slower and larger particles, which resided in the influence of these binary dynamics for longer periods of time, as opposed to the faster and smaller particles in the outer un-curved spirals. These spirals are visible in both NFM and WFM observations, although, they are most prominent in the NFM images from 28 and 29 September. The spirals visibly detached from the system center and moved in an anti-Sunward trajectory, due to acceleration by SRP and trailing, from 01 October onward in NFM imagery. We refer to these structures as wings, which propagated outwards from the system until final detection on 19 October. The wings are denoted by dotted green lines in panels (j) through (n) in Figure \ref{fig:WFM} and (a) through (e) in Figure \ref{fig:WFM_offset}. Below, we report position angle (angles measured from N to E), velocity measurements, and grain size estimations for the aforementioned structures. 

Using our unenhanced polar images, we measured position angles of the northern and southern edges from radial brightness profiles within the first 15$\arcsec$ from the system for WFM and 2$\arcsec$ from the system for NFM. We summed the brightness profiles in the specified region and fit a Gaussian distribution to them, through a similar process to what is shown in \ref{fig:model_fit}. We retrieved the location of each Gaussian peak for both edges, which corresponded to the azimuthal location of the edge. Following this methodology, we found that the average position angles on 27 September, in NFM, are 6.5$\pm$0.9$^{\circ}$ and 134$\pm$0.5$^{\circ}$ for the northern and southern edges, respectively, with a projected opening angle of 127$\pm$1.1$^{\circ}$. Similarly, in WFM we measured position angles of 0.7$\pm$0.5$^{\circ}$ and 132$\pm$0.7$^{\circ}$ for the northern and southern edges, respectively, with an opening angle of 131$\pm$0.8$^{\circ}$. These values are consistent with similar findings reported using the HST \citep{Li:2023}, which operated at similar spatial resolutions to our NFM imagery. 

Next, we sought to characterize the anti-Sunward projected velocity of the wings. We chose to conduct this analysis on the northern wing only, since it was the most visible in our WFM observations. We attempted the same analysis on the southern wing. We could not consistently track the wing over the dates due to low signal to noise, so we did not proceed with the velocity derivation and do not report on this structure. Considering the northern wing maintained a sharp Sunward boundary and its near perpendicular orientation with respect to the tail, we used the corner between the northern wing and tail as a position locator. We estimated the position of the corner on 03 and 19 October in our offset imagery (Fig. \ref{fig:WFM_offset}). We used the change in angular location, time, and the small angle approximation to estimate a velocity of roughly v$_\mathrm{wing}$ = 0.85$\pm$0.45 m s$^{-1}$ for the base of the northern wing. We then used this velocity to estimate the grain size of the northern wing base assuming that the grains are only accelerated by SRP, which may neglect any velocity contributions by gravitation forces or inter-particle collisions. We do not take into account the effects of phase angle in our velocity derivation, so we only report a sky projected velocity. This is true for all other velocities reported in this paper. Assuming dust grains with constant density of $\rho_\mathrm{Didymos}$ = 2170 kg m$^{-3}$ \citep{Naidu:2020}, P$_\mathrm{sun}$($R_{Didymos}$,1.014 au) = 1301 W m$^{-2}$, spherical geometries with radii ranging between r$_\mathrm{min}$ = 1 $\mu$m to r$_\mathrm{max}$ = 1 mm, and perfect scattering, we calculated five distributions of grain velocities corresponding to date of wing detachment, between 01, 03, 04, 07, and 14 October. The equations below show our process:

    \begin{center}
        a$_\mathrm{SRP}$=$\frac{2 P_\mathrm{sun}(R_\mathrm{Didymos}) \pi r_\mathrm{grain}^2 c^{-1}}{\rho_\mathrm{Didymos} V_\mathrm{grain}}$
    \end{center}
    
\begin{figure}[ht!]
        \includegraphics[width=\linewidth]{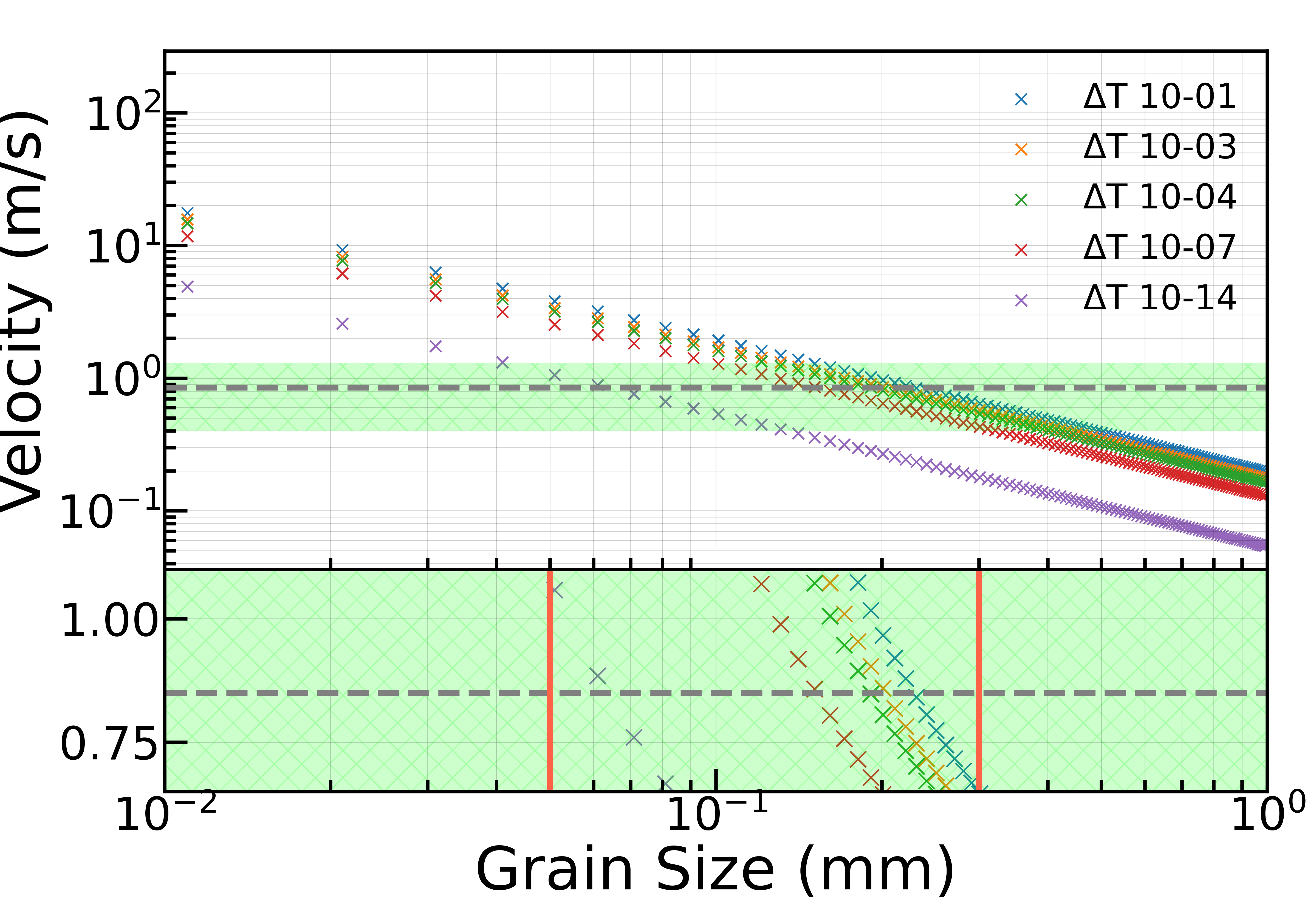}
        \caption{Grain size distribution as a function of wing detachment date, shown as $\Delta$T in legend. The grey dashed line represents the measured Northern wing velocity, with the associated uncertainties shaded around it. The upper panel depicts the entire distribution, while the lower panel offers a zoomed-in window to more closely analyse the size distribution. We derive the bounds on grain size through locating the leftmost and rightmost markers that occupy the uncertainty threshold. The leftmost and rightmost markers are associated with the smallest and largest radii, respectively, between 0.05 and 0.2mm, denoted by the red vertical lines. \label{fig:wings}}
    \end{figure}

    \begin{center}
        $v_\mathrm{particle}$ = a$_\mathrm{SRP}$ $\Delta$T
    \end{center}

where a$_\mathrm{SRP}$ is the acceleration on the particles due to SRP, r$_\mathrm{grain}$ denotes the value between r$_\mathrm{min}$ and r$_\mathrm{max}$, V$_\mathrm{grain}$ is the computed grain volume, and $\Delta$T is the time between spiral detachment and last detection on 19 October. We compared the distributions of computed grain velocities with the measured northern wing velocity, which is represented by the dashed line and surrounded by a green uncertainty threshold in the upper panel of Figure \ref{fig:wings}. We identified where the distributions crossed through the wing velocity threshold, and used the intersections to put upper and lower bounds on grain size at the base of the northern wing's Sunward-edge. Finally, from figure \ref{fig:wings}, we infer that the Northern wing base is populated by particles in the range of approximately 0.05 to 0.2 mm, which are likely the largest particles in the wing and are consistent with upper grain size bounds in the primary tail between 01 and 15 October retrieved by \cite{Li:2023}. An important distinction between our methodology and that of \cite{Li:2023} is that we used the bulk density of Dimorphos as the density of the grains, which accounts for porosity inside the grains, while \cite{Li:2023} used an averaged chondritic grain density of around 3500 kg m$^{-3}$. Despite this difference, our derived grain sizes are consistent and well within each others uncertainties.

    \begin{table*}[h!]
        \caption{Ejecta Cone Position and Opening Angles}
        \centering
        \begin{tabular}{lcccccccc}
        \hline\hline
        Date & Time (UT) & Mode & N-Edge ($^{\circ}$) & Avg ($^{\circ}$) & S-Edge ($^{\circ}$) & Avg ($^{\circ}$) & Opening Angle ($^{\circ}$) & Avg ($^{\circ}$) \\
        \hline
        22-09-27 & 05:36:21 & NFM & 5$\pm$1.7 & | & 136$\pm$0.7 & | & 129$\pm$1.9 & |   \\
        22-09-27 & 07:48:36 & NFM & 8$\pm$0.8 & 6.5$\pm$0.9 & 132$\pm$0.8 & 134$\pm$0.5 & 124$\pm$1.1 & 127$\pm$1.1 \\
        22-09-28 & 06:25:11 & NFM & 13$\pm$0.9 & X & 119$\pm$1.3 & X & 105$\pm$1.6 & X \\
        \hline
        22-09-27 & 03:21:05 & WFM & 1.2$\pm$0.4 & | & 133$\pm$1.7 & | & 132$\pm$1.7 & |   \\
        22-09-27 & 04:25:31 & WFM & 0.8$\pm$1.3 & | & 130$\pm$1.0 & | & 129$\pm$1.6 & |   \\
        22-09-27 & 06:58:05 & WFM & 0.1$\pm$0.2 & 0.7$\pm$0.5 & 132$\pm$0.4 & 132$\pm$0.7 & 132$\pm$0.5 & 131$\pm$0.8 \\
        22-09-28 & 04:28:06 & WFM & 3.6$\pm$0.4 & | & 131$\pm$1.0 & | & 128$\pm$1.1 & |   \\
        22-09-28 & 05:13:29 & WFM & 0.9$\pm$1.3 & | & 131$\pm$0.1 & | & 130$\pm$1.3 & |   \\
        22-09-28 & 07:52:24 & WFM & 3.6$\pm$0.4 & 2.7$\pm$0.5 & 131$\pm$1.0 & 131$\pm$0.5 & 128$\pm$1.1 & 129$\pm$0.7 \\
        \hline        
        \end{tabular}
        \newline
        \newline
        \textbf{Notes.}{ Ejecta cone position angle measurements from NFM and WFM imagery. N-egde and S-edge denote the retrieved position angles of the edges per image per night. Opening angle is the difference between those edges, with propagated uncertainties applied, per image per night. Avg columns denote the average value for N-edge, S-edge, and Opening Angle per night. X in the Avg column indicates that there was only one image taken that night and averaging need not be applied.}
        \label{tab:cone}
    \end{table*}

    \begin{table*}[h!]
        \caption{Wide Field Mode Clump Velocities}
        \centering
        \begin{tabular}{lcccc}
        \hline\hline
        Distance & Clump ID & Velocity$_\mathrm{AB}$ (m s$^{-1}$) & Velocity$_\mathrm{AD}$ (m s$^{-1}$) & Velocity$_\mathrm{BD}$ (m s$^{-1}$) \\
        \hline
        September 27: & 04:25:31-06:58:05 UT &  &  &   \\
        \hline
        13.5$\arcsec$ & C1-S & 10.1$\pm$6.5 & 11.4$\pm$2.0 &  11.0$\pm$2.3 \\
        11.0$\arcsec$ & C1-N & 8.1$\pm$6.8 & 9.3$\pm$1.3 &  8.9$\pm$2.7 \\
        8.0$\arcsec$ & C2-S & 6.3$\pm$3.9 & 6.3$\pm$1.3 &  6.3$\pm$1.4 \\
        4.8$\arcsec$ & C3.1-N & 2.8$\pm$3.6 & 3.9$\pm$1.1 &  3.5$\pm$1.6 \\
        3.5$\arcsec$ & C3-S & 2.9$\pm$5.3 & 2.6$\pm$1.2 &  2.6$\pm$1.7 \\
        2.7$\arcsec$ & C3.2-N & 2.0$\pm$5.5 & 2.2$\pm$2.4 &  2.1$\pm$2.2 \\
        \hline
        September 28:  & 04:28:06-07:52:24 UT &  &  &   \\
        \hline
        20.5$\arcsec$ & C4-N & 4.1$\pm$3.7 & 3.5$\pm$0.3 &  3.6$\pm$0.5 \\
        19.2$\arcsec$ & C4-S & 2.5$\pm$3.9 & 3.2$\pm$0.6 &  3.2$\pm$0.4 \\
        11.5$\arcsec$ & C5-N & 1.5$\pm$4.3 & 1.9$\pm$0.7 &  1.9$\pm$0.6 \\
        7.9$\arcsec$ & C5-S & 1.3$\pm$3.9 & 1.3$\pm$0.5 &  1.3$\pm$0.4 \\
        \hline     
        \end{tabular}
        \newline
        \newline
        \textbf{Notes.}{ Derived velocities for clumps in WFM. Distance is the radial distance at which the final detection occurred. Clump ID is a way to efficiently refer to the clumps in text, roughly paired in bins of similar radial distance and book-ended by the suffix or N or S to denote position in Northern or Southern region of the cone. Additionally, if two clumps are from the same region of the cone and in the same radial bin, they receive an additional decimal marker, with 0.1 denoting the outermost of the clumps while 0.2 denotes the innermost. Velocity$_\mathrm{AB}$ is the calculated velocity between appearances of the clump, Velocity$_\mathrm{AD}$ is the velocity between first appearance and Didymos, and Velocity$_\mathrm{BD}$ is velocity between second appearance and Didymos.}
        \label{tab:clump_WFM}
    \end{table*}

    \begin{table*}[h!]
        \caption{Narrow Field Mode Clump Velocities}
        \centering
        \begin{tabular}{lcccc}
        \hline\hline
        Distance & Clump ID & Velocity$_\mathrm{AB}$ (m s$^{-1}$) & Velocity$_\mathrm{DA}$ (m s$^{-1}$) & Velocity$_\mathrm{DB}$ (m s$^{-1}$) \\
        \hline
        28 - 29 September: & 06:25:11-08:28:06 UT &  &  &   \\
        \hline
        1.3$\arcsec$ & C6-N & 0.05$\pm$0.08 & 0.19$\pm$0.08 &  0.13$\pm$0.05 \\
        \hline
        28 September - 01 October:  & 06:25:11-07:03:01 UT &  &  &   \\
        \hline
        1.4$\arcsec$ & C6.1-S & 0.10$\pm$0.03 & 0.13$\pm$0.08 &  0.11$\pm$0.03 \\
        1.2$\arcsec$ & C6.2-S & 0.09$\pm$0.03 & 0.11$\pm$0.08 &  0.09$\pm$0.03 \\
        \hline 
        \end{tabular}
        \newline
        \newline
        \textbf{Notes.}{ Derived velocities for clumps in NFM. Velocity$_\mathrm{AB}$ is the calculated velocity between appearances of the clump, Velocity$_\mathrm{AD}$ is the velocity between first appearance and Didymos, and Velocity$_\mathrm{BD}$ is velocity between second appearance and Didymos.}
        \label{tab:clump_NFM}
    \end{table*}

\subsubsection{Ejecta Cone Clumps} \label{subsubsec:clumps}

Various other features were observed in the expanding ejecta cone and debris cloud, such as bright regions called clumps. Clumps are likely either increased local grain densities, line of sight enhancements from optical depth, or physical structures comprised of larger grains. We identified clumps by first isolating series of unenhanced Cartesian images across a single night and observing mode. We then searched for bright regions in those images that were spatially separated throughout the FoV, and subsequently catalogued those bright regions. We approximated the center of these regions and cross-checked with the enhanced Cartesian images for a similar structure at the estimated center. Using this methodology, we identified and tracked 13 clumps at various radial distances from the binary, with the majority of clumps detections occurring within the first two days post-impact using WFM images. Only three clumps were identified after 28 September, which were exclusively observed in NFM images. For our WFM clumps, we then input the approximated clump centers into a 2D Gaussian centroiding algorithm, which precisely located the peak of each clump within an 11x11 pixel search area that corresponded to 2 $\times$ the average seeing conditions for each night, as seen by the white boxes in both panels of Figure \ref{fig:velocity}. For the NFM clumps, we did not increase the 11x11 pixel search area to cover 2 $\times$ the average seeing, since NFM observations benefited from AO and were therefore less affected by atmospheric dispersion. Additionally, the quality of our NFM observations were on the order of 100 mAs resolution, which corresponds to the 11x11, or 130x130 mAs, search area. Throughout the following figures, Figs \ref{fig:WFM_clump} and \ref{fig:NFM_clump}, we present the enhanced Cartesian images corresponding to the last point of detection for each of the clumps. For the WFM clumps, the clumps were solely tracked over a series of exposures from one night of observations. Generally, our NFM clumps were retrieved at later dates, and thus were comprised of slower and larger ejecta. In order to measure their displacement, we needed to use observations that spanned multiple nights. Each clump is associated with their own cutout, where the green marker represents the calculated 2D Gaussian centroid of each clump. Clumps IDs can be found in the corners of these cutouts. 

We computed the sky projected velocity of the clumps in three different ways; (i) from first detection to last detection, (ii) from first detection to Didymos-Dimorphos barycenter, and (iii) from last detection to Didymos-Dimorphos barycenter. We used method (i) to gauge the relative clump velocity between images to retroactively extrapolate clump position and determine the time and place of origin of a clump. After we verified that a clump originated from the Didymos system around the time of impact, we then used methods (ii) and (iii) to calculate clump velocities across a larger spatial and temporal baseline, which minimized our uncertainties. Discrepancies remained between velocities measured using methods (ii) and (iii), and are likely due to either changing observational conditions or internal morphological change within the clump that slightly shifted centroids between first and last detection. We report the derived plane-of-sky clump velocities and associated uncertainties in Tables \ref{tab:clump_WFM} and \ref{tab:clump_NFM} for WFM and NFM, respectively. As seen in Table \ref{tab:clump_WFM}, there is a strong correlation between clump velocity, radial distance, and date of detection. The earliest group of clumps were sampled from T+0.15D - T+0.28D, and exhibit a positive correlation between velocity and radial distance from the system center. Such phenomena can likely be attributed to the nature of the impact as an impulsive event, where all ejecta is assumed to have been ejected at the time of impact and smaller less massive particles had higher ejection speeds than larger more massive particles. Therefore, outermost clumps should have higher velocities and smaller particles, while innermost clumps have the lowest velocities and largest particles. Our lowest calculated velocity is associated with the latest and innermost retrieved clump, exhibiting a velocity around 0.09$\pm$0.03 m s$^{-1}$ for NFM clump C6.2-S from 28 September to 01 October, at an initial distance of 1.2$\arcsec$ or 63.8 km from the binary. These slowest and closest clumps, specifically in the southern spiral in NFM imagery, also exhibited some non-radial movement in a clockwise direction, which is similar to the overall curvature of the southern spiral. 

    \begin{figure*}[p!]
    \begin{center}
        \includegraphics[scale=0.62]{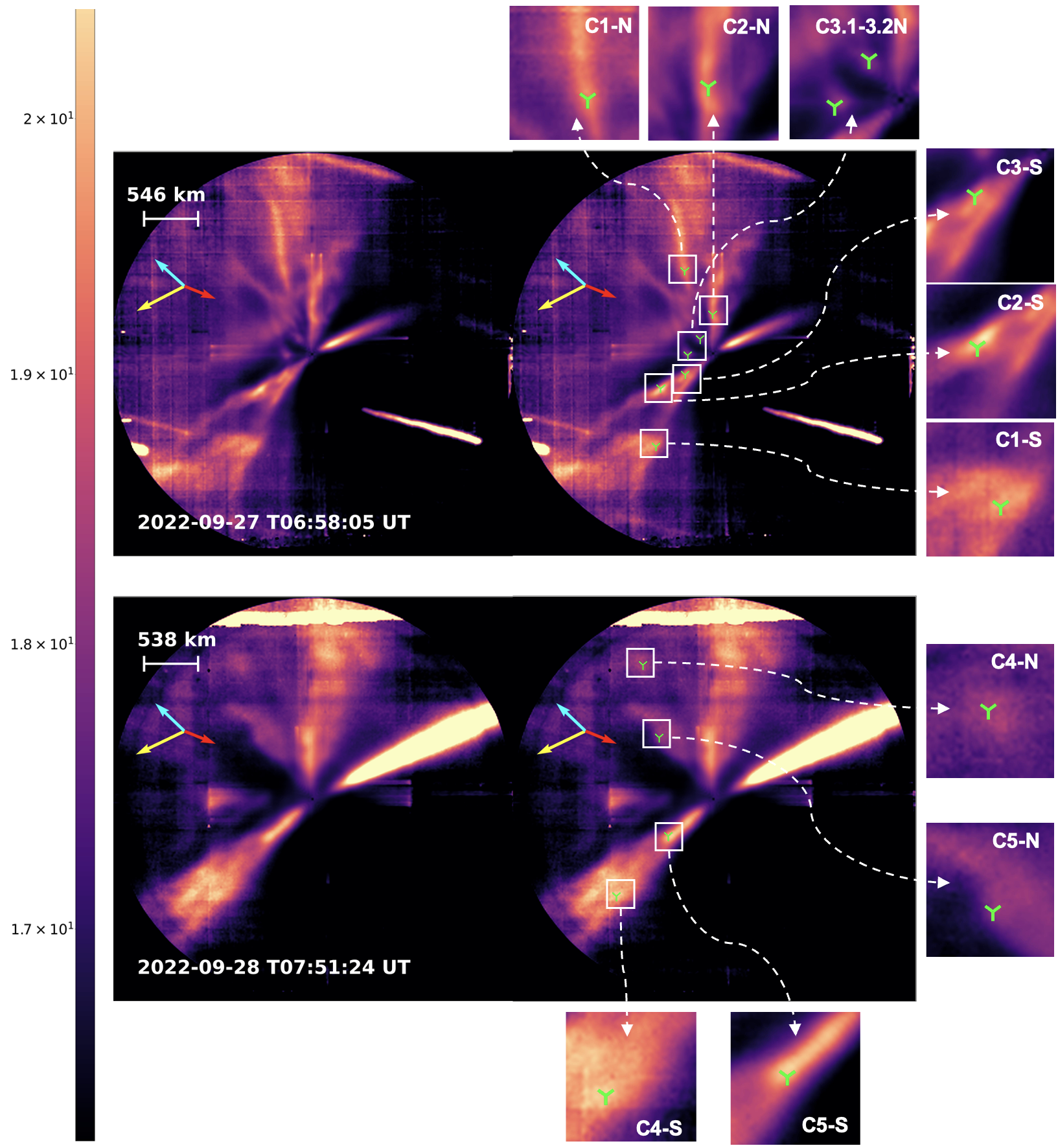}
        \caption{Illustration of the ejecta clumps identified within the first 48hrs post impact in WFM (200-2500km from system center). The images are based on the last detection of the clump, which allows for greater clarity due to the higher radial distance from the system. The green markers denote the computed centroid for each clump. Intensities are in units of 10$^{-20}$ erg/s/cm$^2$. The yellow arrow indicates sun-angle, cyan arrow represents heliocentric orbital motion of the system, and red arrow is the DART spacecraft approach angle.  \label{fig:WFM_clump}}
    \end{center}
    \end{figure*}

    \begin{figure*}[p!]
    \begin{center}
        \includegraphics[scale=0.62]{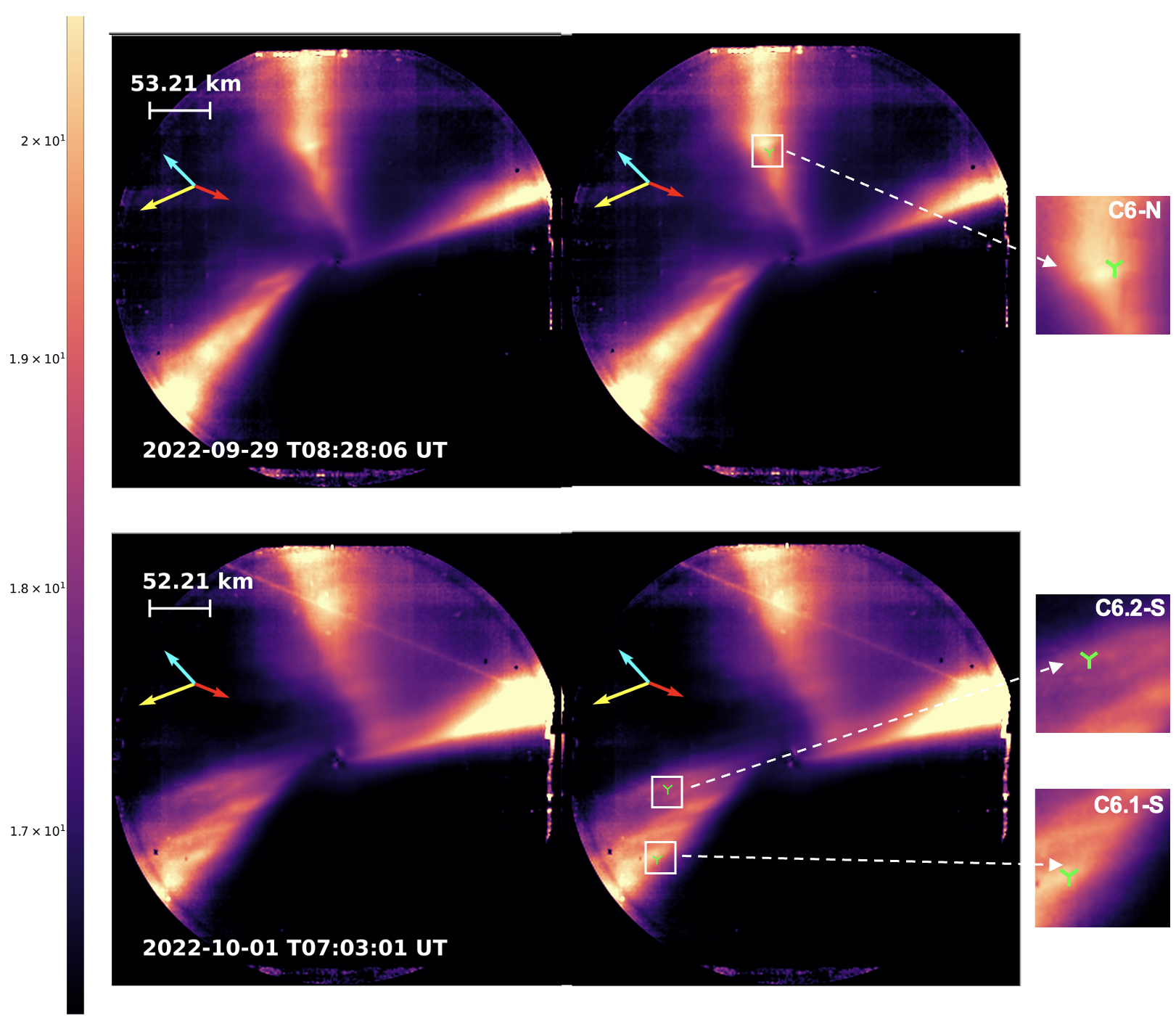}
        \caption{Illustration of the ejecta clumps identified within the first 6 days post impact in NFM (20-220km from system center). The images are based on the last detection of the clump, which allows for greater clarity due to the higher radial distance from the system. The green markers denote the computed centroid for each clump. Intensities are in units of 10$^{-20}$ erg/s/cm2. The yellow arrow indicates sun-angle, cyan arrow represents heliocentric orbital motion of the system, and red arrow is the DART spacecraft approach angle. \label{fig:NFM_clump}}
    \end{center}
    \end{figure*}

    \begin{figure}[h]
        \includegraphics[width=\linewidth]{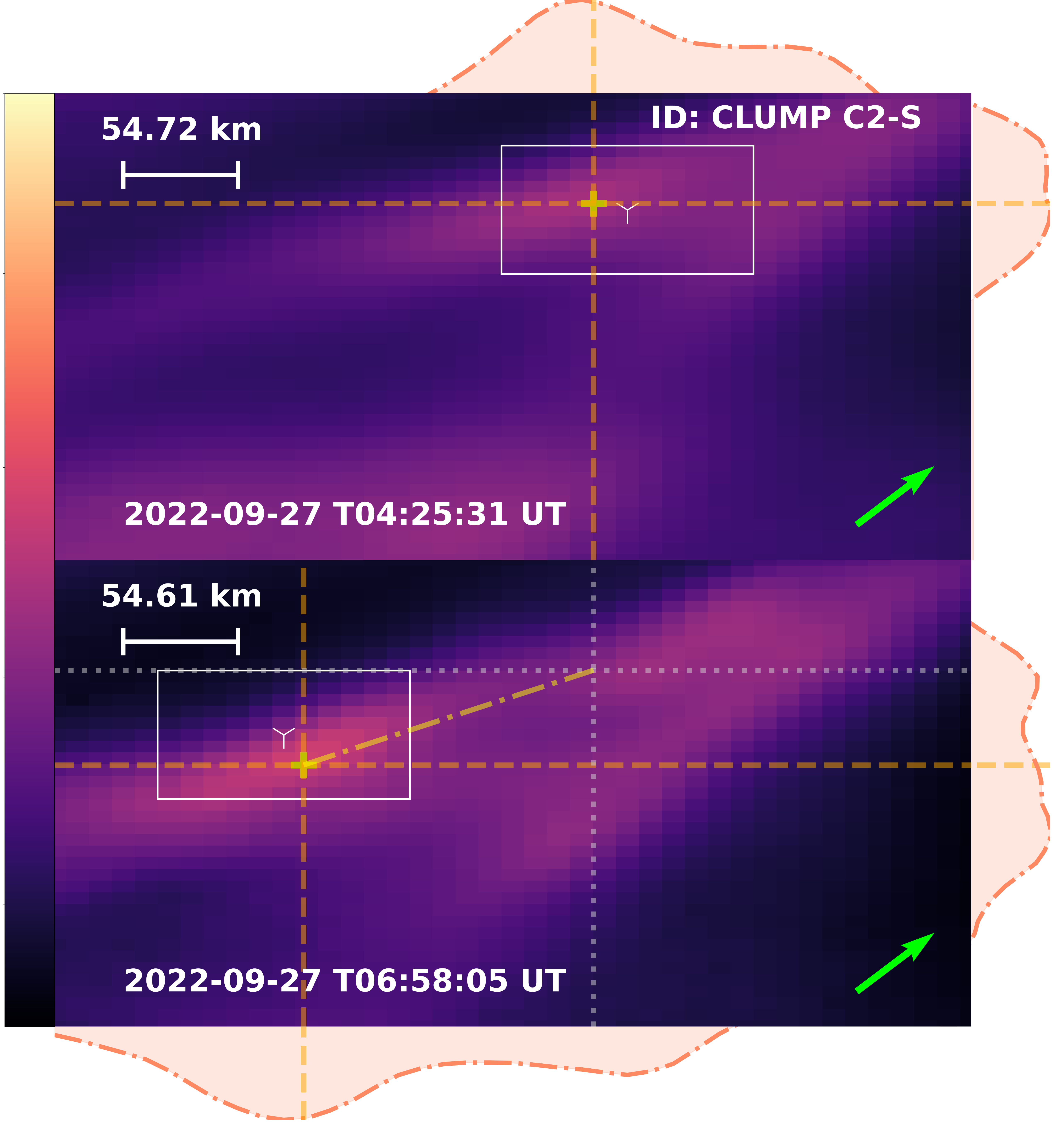}
        \caption{Velocity of clump C2-S, first seen in the second MUSE WFM exposure in the southern ejecta cone edge at T03:21:05 UT 2022-09-27, tracked from third to fourth exposures. In panels (a) and (b) orange lines denote the axes of the calculated centroid, which is depicted by a yellow cross. Panel (b) has grey and yellow lines, which show the previous position of the clump centroid and the approximate path it took to reach its current position. Both panels have marginal density plots to highlight the structure of the clump as sliced by the orange lines in each image. Intensities are in units of 10$^{-20}$ erg/s/cm$^2$. The green arrow is the angle toward Didymos center, and the white scale bar is distance per 1 arcsecond.\label{fig:velocity}}
    \end{figure}

    \begin{figure*}
        \plotone{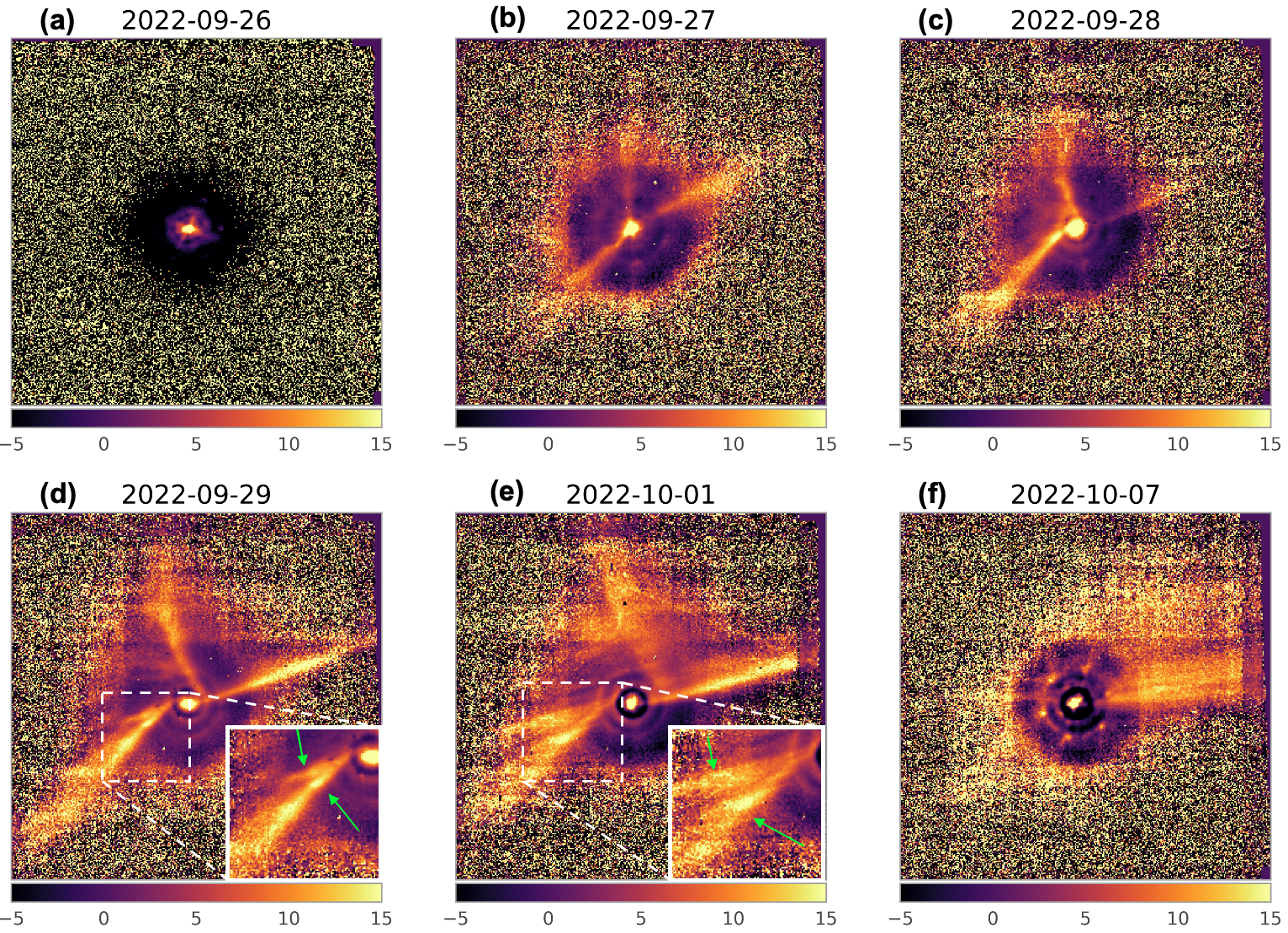}
        \caption{Maps of relative reflectance from NFM observations, spanning 500-750 nm and normalized at 500 nm due to the AO lasers. The maps are oriented with North up and East left, the date above each map is the average date of the stack, and the reflectance slope indicated by the colour bar is expressed in units of $\%$/100 nm. The dark annuli around the sources are artefacts from our analysis, since shorter and longer wavelengths create different point spread functions around the central condensation, which in turn created regions of negative reflectance slope (black rings). We therefore do not compare the colours of the central condensation. Clumps C6.1S and C6.2S are highlighted by the inset zoomed region, with lime arrows marking each clump.
        \label{fig:color}}
    \end{figure*}

Simultaneous to our analysis of the images and similar to what is done in \cite{Opitom:2023}, we analysed NFM reflectance maps that characterized the dust environment up to around 220 km from the system center. Generally, changes in the reflectance maps are indicative of changes in the properties of the dust particles, such as difference in composition or size \citep{Jewitt:1986}. In our analysis, we assumed change in dust particle size drove color differences in the reflectance maps, since the majority of observed particles likely originated from Dimorphos. Typically, shallower reflectance slopes are associated with bluer colors and thus smaller particles, while steeper slopes are indicative of redder and larger particles \citep{Jewitt:1986}. These color changes are illustrated through panels (a) to (e) in Figure \ref{fig:color}, where three main features associated with the northern spiral, southern spiral, and tail appear between 26 and 27 September with slopes around $5-6\%/100$~nm and consistently steepen/redden into 28 September. By 29 September, the reflectance slopes of all three structures approached $8-10\%/100$~nm, with notable regions of higher reflectance slopes. Three of these steeper slope regions spatially coincide with observed clumps in NFM images. We measured reflectance slopes in each of these features on the order of $13-15\%/100$~nm, which is almost a factor of two higher than the surrounding dust reflectance slopes. Since these clumps exhibited higher reflectance slopes we postulate that they are likely comprised of larger dust particles than the surrounding \textbf{dust}. Velocity measurements further provide evidence for this claim, since the velocities of these redder clumps are significantly slower than all other clumps. Furthermore, we see that these redder clumps also travel non-radially with the southern spiral, which reinforces them being comprised of larger more massive particles that were lower velocity and thus had more time to be influenced by the binary gravitational forces that also influenced the spirals.

\subsection{Tail Structure and Evolution} \label{subsec:tail}

As described by \cite{Li:2023, Lin:2023, Moreno:2022}, following the formation of the ejecta cone, the initial micron to sub-millimeter ejecta was quickly scattered by gravitational interactions with the binary and rapidly accelerated out of the system by SRP, which formed the primary dust tail first observed in WFM at T+0.22D after impact.  Intriguingly, we observed in WFM the formation of a secondary tail around 03 October (T+6.4D) after impact, which manifested as a positive skew in brightness profiles of the primary tail until it eventually became fully resolved by 14 October (T+17.3D) after impact. In the following sections, we discuss the primary and secondary tails and report our characterization of both structures.

    \begin{figure*}[ht!]
        \includegraphics[width=\linewidth]{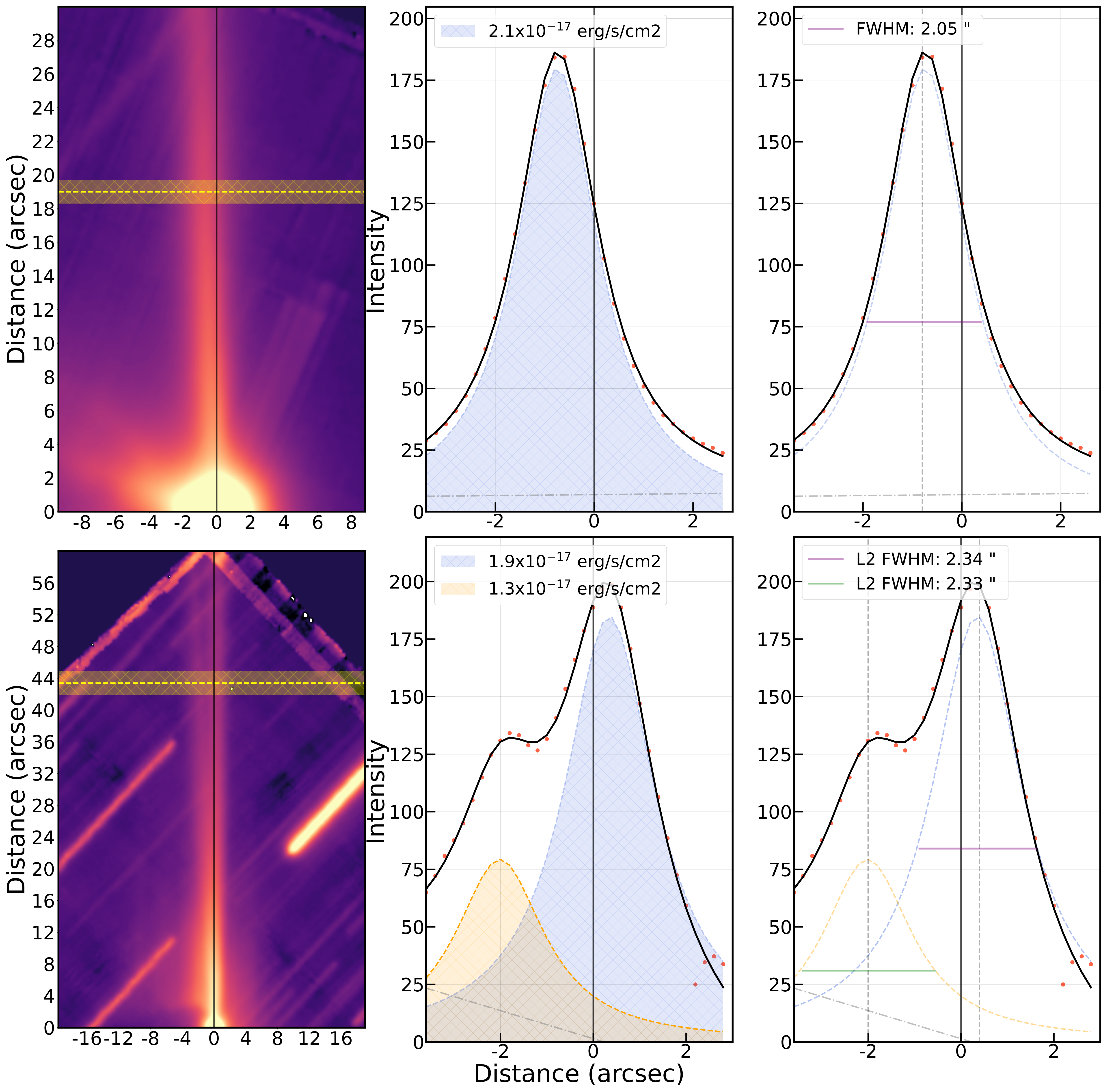}
        \caption{Model fitting process for the primary tail (upper panels) and dual tails (lower panels), from 03 October T08:22:08 UT and 14 October T07:35:31 UT, respectively. The left-most panels show the derotated Cartesian image, with a black solid line denoting the average position angle of the tail, a yellow dashed line for the radial midpoint of the N-summed profile, and yellow shaded area for the range rows summed to make the profile. The middle panels depict the fit components (dashed blue is primary tail and dashed orange is secondary tail), background contribution (dot-dash grey), best fit model (solid black), original data points (red circles), and integrated flux (shaded areas). The right most panels demonstrate the fit components again, the central axis of each component (vertical grey dashed line), and the FWHM measurements (purple for primary tail and green for secondary tail).  \label{fig:model_fit}}
    \end{figure*}

We built a pipeline to consistently measure tail position angle, width, and flux as a function of radial distance from Didymos. In this pipeline, we fitted Lorentzian distributions which were allowed to vary over amplitude (a), characteristic width ($\sigma$), and center value ($\mu$) for a given brightness profile taken across the tail. These brightness profiles were extracted row by row from our unenhanced polar images, which were discussed in Section \ref{sec:image proc}. Each modeled brightness profile was a summation of $\pm$N/2 rows above and below a centered radial midpoint, where N generally equalled 2 $\times$ the average atmospheric seeing of the night. We summed the profiles in this way to increase the signal to noise in our fits and not over-sample the tail while accounting for atmospheric dispersion in our observations. We subsequently increased N to 3-4 $\times$ atmospheric seeing from 07 October onward due to the system becoming fainter due to intrinsic fading post-impact, increasing geocentric distance, and the increasingly crowded background field as Didymos crossed the galactic equator. However, certain fits remained unusable due to passing background stars, so they were rejected. 

We used a 2 stage approach for the fitting process. We started with an initial fit of the first 10$\arcsec$ of summed brightness profiles to refine initial estimations of the input fit components, and then applied fits with these updated components to all N-summed profiles. We utilized this method to extract position angles and their respective uncertainties from all observations where the primary tail was fully formed, roughly our entire observational campaign. Notably, polar images are ideal for determining angular position data, however, they do not conserve the flux and spatial portion of the original data. Subsequently, in the second step, we used the pipeline to derotate Cartesian images by the average position angle of the tail so the tail is oriented vertically, which allowed for us to easily sample, sum, and extract tail widths and fluxes. Finally, we used these fits to estimate the widths, fluxes, and their associated uncertainties along the same timescale as the position angles. Previous studies of other active asteroid tails have used similar methods for tail characterization, such as that of (3200) Phaethon's recent dust tail \citep{Battams:2022}.

In the case of the secondary tail, we modified both pipelines to support dual Lorentzian fitting to separate the dual tails in the skewed profile from 03 to 07 October, and fully model the two on 14 October. Figure \ref{fig:model_fit} demonstrates our process for fitting and measuring the brightness profiles, specifically the dual Lorentzian fit of both tails in the lower portion of the panel.

\subsubsection{Primary Tail Evolution} \label{subsubsec:primary}

\begin{figure}[ht!]
        \includegraphics[width=\linewidth]{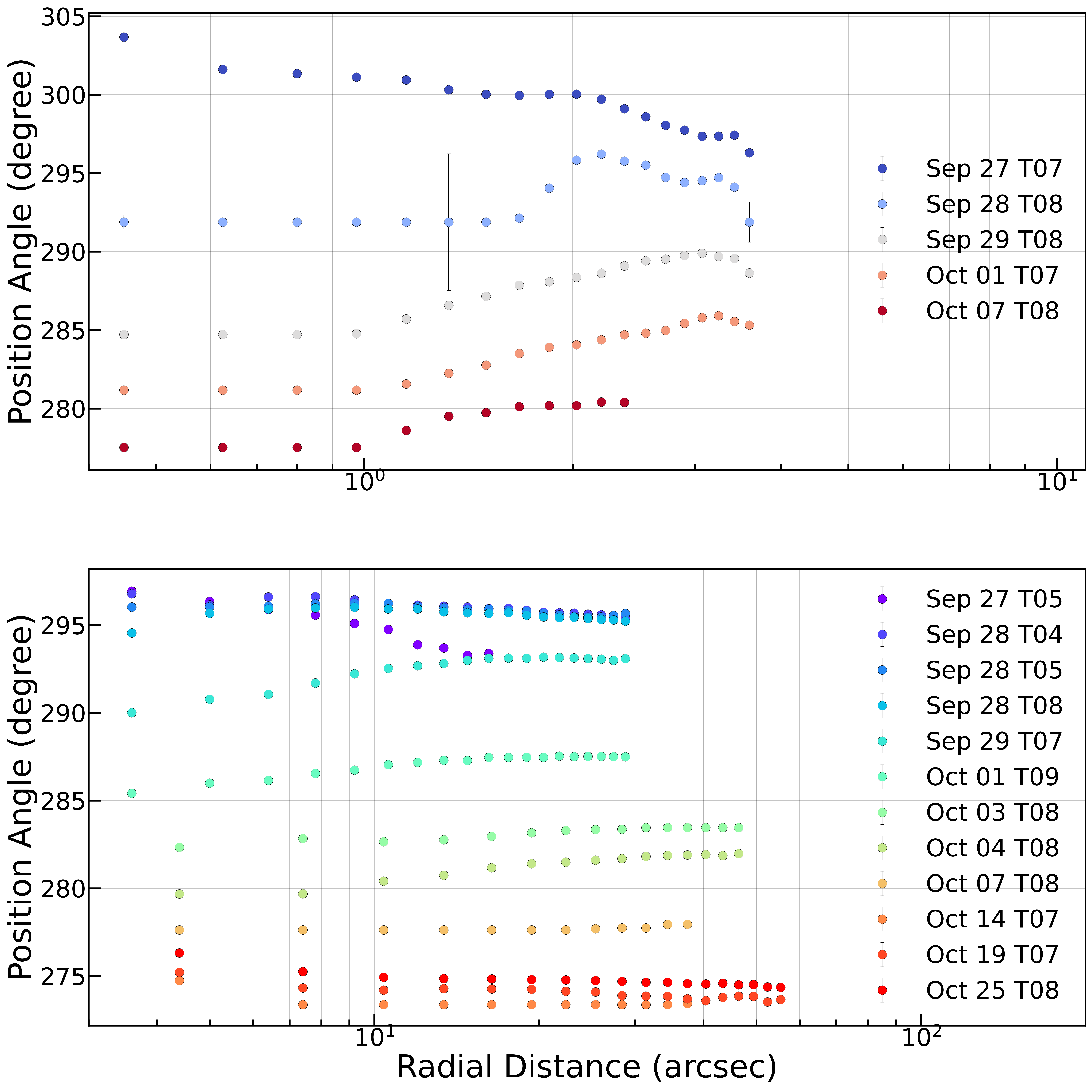}
        \caption{Primary tail position angles as measured from North to East from the Lorentzian fits. Cooler colours represent earlier dates in the campaign, while warmer colours are later dates. Uncertainties associated with the measurements are present, however, are smaller generally than the marker-size of the plot.\label{fig:t1_pa}}
    \end{figure}

The properties of the primary tail changed over time, because of viewing geometry, initial ejecta velocities, grain size differentiation, and binary gravitational dynamics. Figure \ref{fig:t1_pa} depicts how the position angle changed over distance and time. Similar to the spirals, the tail exhibited significant curvature within the first 5-10$\arcsec$ from the system, from 27 September until 01 October. Notably, this curvature is prominent within the first week after impact and manifests in both clockwise and counter-clockwise rotation, with a rotation inflection point on 28 September for both observing modes. We can also see that prior to 28 September, the position angles decrease with distance, while after 28 September they increase with distance until 07 October. After 07 October, the position angles return to a slight decrease with distance, marking another inflection point. Intriguingly, a large jump in position angles is shown at 2$\arcsec$ on 28 September, the date of the first slope and curvature inflection. The nature of this bump and possible relation to the inflections are not yet known. We speculate that the jump could possibly be a result of binary gravitational interactions with slower grains at the base of the tail, which had originally dominated the shape of the tail within a few arcseconds of the source as shown in models by \cite{Ferrari:2022}. We also note the role of changing viewing geometry, which may explain the position angle slope inversion around 04 to 07 October in our WFM measurements, since the orbital plane angle and geocentric distances were changing during this time. The magnitude of the change in orbital plane angle is a few degrees, which is consistent to the change in slope of the position angles, which may indicate a connection. Similar position angles and curvature have also been reported by \cite{Li:2023}, however, minor discrepancies remain between our work that are likely due to differing viewing geometry between HST and MUSE.

    \begin{figure}[ht!]
        \includegraphics[width=\linewidth]{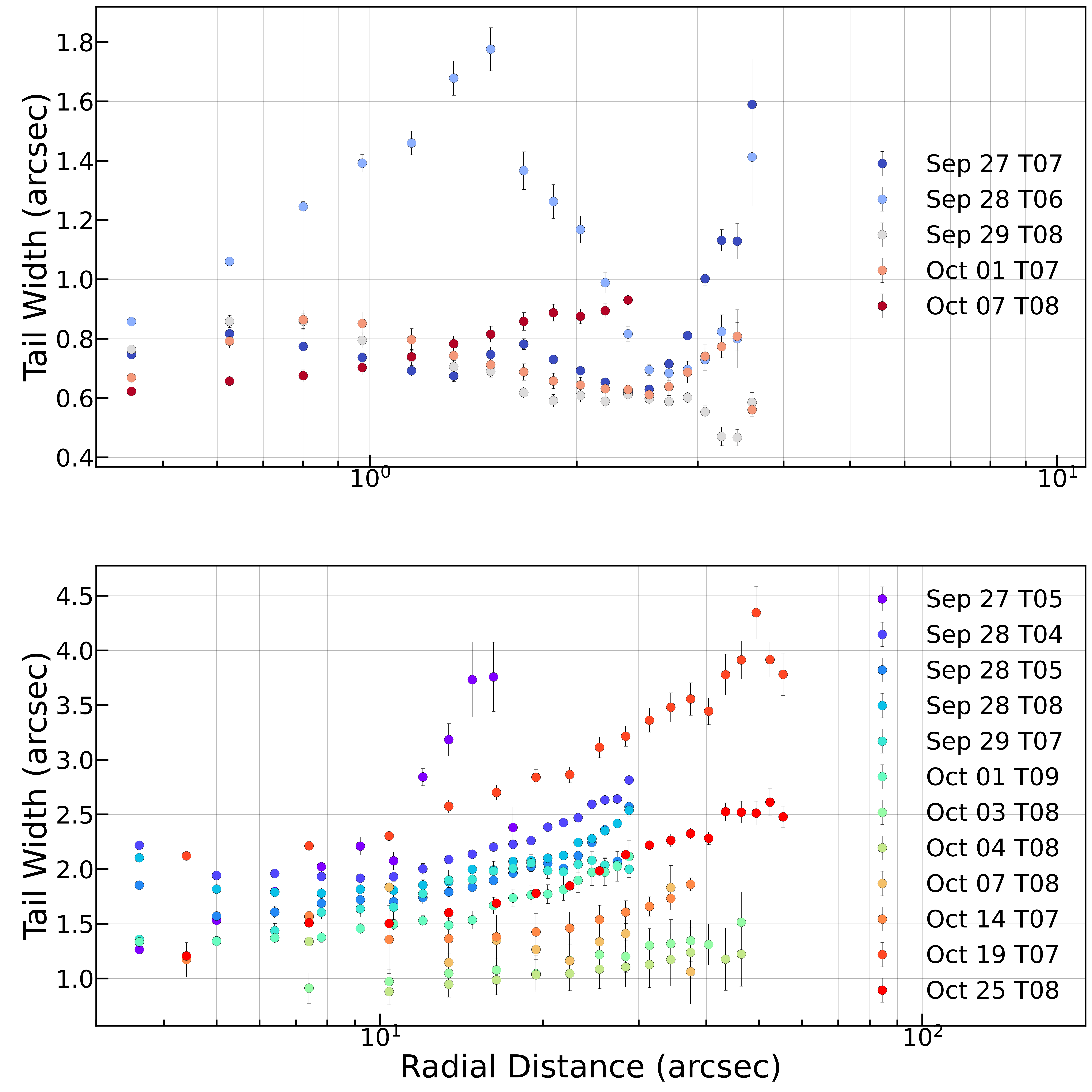}
        \caption{Primary tail widths as measured from the FWHM of the Lorentzian fits. Cooler colours represent earlier dates in the campaign, while warmer colours are later dates. Width measurements on 19 October cannot be commented on, as atmospheric seeing for that night is on the order of 3-5 $\times$ higher than all other observations, and have positively biased our measurements.\label{fig:t1_width}}
    \end{figure}

Width measurements are more sensitive to atmospheric seeing conditions, and therefore have higher intrinsic variance between values and greater uncertainties. We measured primary tail widths within the first 200km of the binary on the order of 0.7-1.1$\arcsec$ in NFM, which are in agreement with other findings at similar spatial resolutions by HST \citep{Li:2023}. Larger widths from 1-4$\arcsec$ were measured in WFM at distances around 200-2500km, with increasing width over increasing distance. We postulate that the outer regions of the tail are comprised of smaller faster particles with higher radial velocity components. Over the timescale it took for those faster particles to reach the outer tail, their radial velocity dispersed them further from the central axis of the tail and thus increased the tail width. Conversely, the inner region of the tail is likely comprised of larger slower particles, which have lower radial velocities and are less dispersed from the central axis. 

Over the course of our campaign, the average primary tail width slowly decreased until 01 October, after which there is a significant decrease for the dates of 03 and 04 October, and slow increase from that minimum width to a maximum width on 19 October. It is important to note, during these later dates, the secondary tail had formed and was co-evolving with the primary, which complicates the interpretation of width measurements. This trend and its interpretation will be discussed in detail throughout Section \ref{subsubsec:secondary}. We also note that atmospheric seeing conditions on 19 October were on the order of 2$\arcsec$, which is a factor of 5 higher than the adjacent nights. Therefore, the width maximum on 19 October is likely a product of observational conditions, and we will not comment further on width measurements from that night. Despite these influences, we also propose that binary dynamics could also have played a role in tail width over time and specifically governed the trajectories and types of particles fed into the tails. Validation of this would require modeling, which is outside the scope of this work. 

    \begin{figure}[ht!]
        \includegraphics[width=\linewidth]{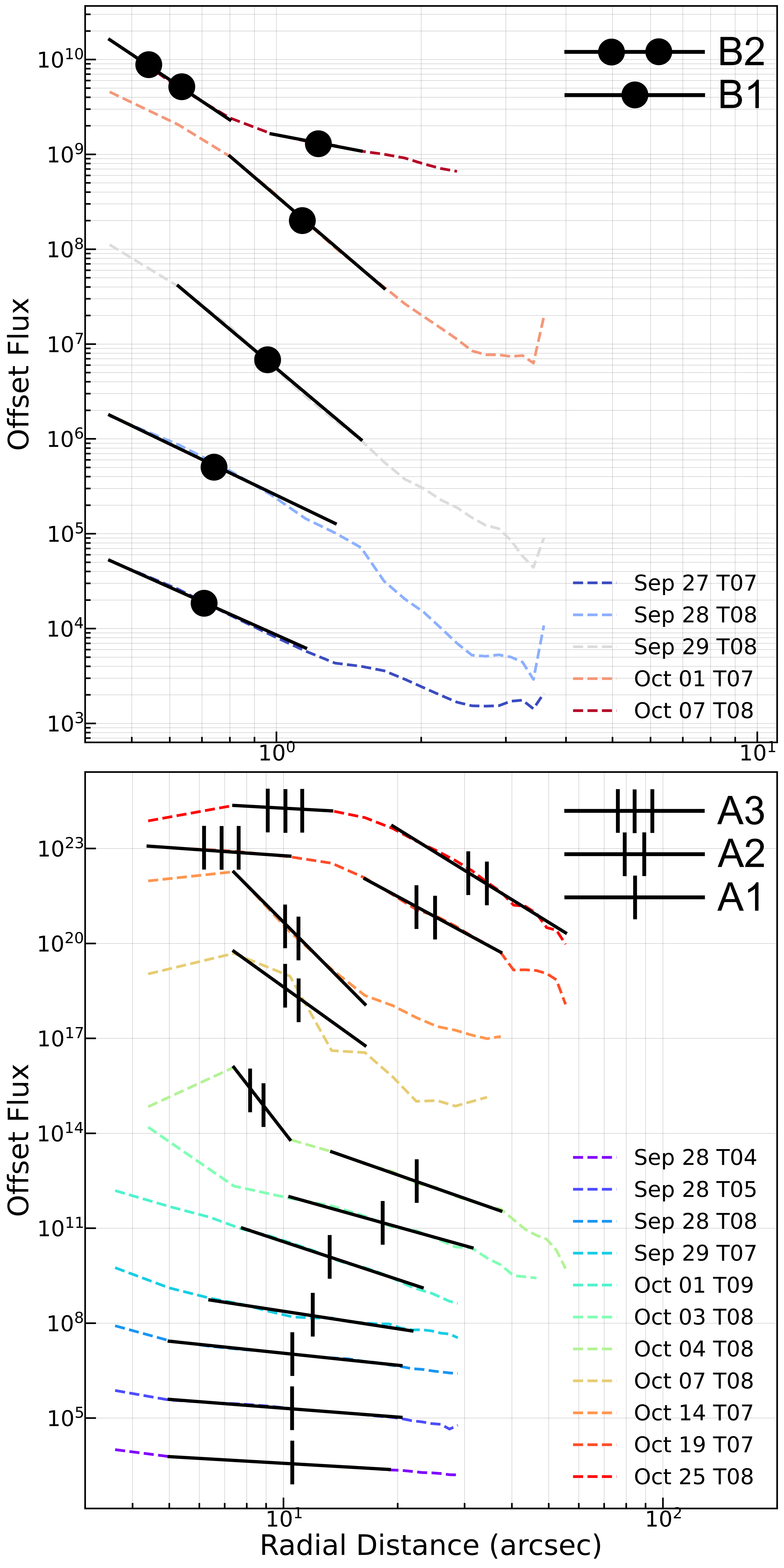}
        \caption{Primary tail relative flux profiles as measured from the integrated area under the Lorentzian fits. Cooler colours represent earlier dates in the campaign, while warmer colours are later dates. Uncertainties have been removed from the measurements, since we have artificially offset the profiles to compare their relative slopes. Black lines depict the regions where we applied power-law fits. \label{fig:t1_fluxshift}}
    \end{figure}

Since MUSE is a spectrograph, we do not have sufficient accuracy in absolute flux calibration of the white light images to compare fluxes directly. Instead, we compared the relative flux intensities along the tail in Figure \ref{fig:t1_fluxshift}. Here, we identified areas of profiles to fit power-laws to, from which we extracted slopes. We identified these regions sequentially, and first established a midpoint of 10 arcseconds from the source in WFM to perform the first fit around. As time progressed, we shifted the midpoint to further radial distances as the relative flux profiles changed in order to maintain sampling on the same region of the tail as it propagated outwards due to SRP. Using this technique, we established three regions of fitted slopes in WFM and two regions in NFM. The first series of power-law fits in WFM, from 28 September to 04 October,  termed A1, span from around 270 to 1100 km from the system, and exhibited a consistently steepening slope from -0.6 to -1.3. Similarly, the second series of fits, A2, ranging from $\sim$ 210 to 1700 km, from 04 to 25 October, also exhibited an initially steepening slope over time, from -2.5 to a maximum of -4.4, but shallows from 14 to 25 October. The final series of fits, A3, ranging from $\sim$ 210 to 590 km from 19 to 25 October, denotes a plateau in brightness, where retrieved slopes shallow further, from -0.2 to -0.1 by 25 October. Our NFM fits depict strikingly similar trends, with the first series of fits (B1, r$\approx$30-155 km, 27 September to 07 October) shallowing from -2.3 to -0.4 and second series (B2, r$\approx$30-55 km, 07 October) with a slope of -1.4. It is clear that the steepening slopes and eventual plateau are indicative of different scattering regimes within the tail, and thus different grain sizes, as argued by \cite{Li:2023}. Furthermore, the slopes and relative flux intensities can be understood as a result of the particle velocity distributions for finite events, such as the DART impact, where smaller less massive particles were impulsively ejected and fed into the tail before larger more massive particles. Therefore, if we assume that the majority of the particles in the tail were from this finite event, they should obey this relation and cause a step-like evolution of the flux intensities. Finally, our measurements from A1 support \cite{Li:2023}'s proposition that the bulk of ejecta is limited to below cm-size grains, which was deduced from their retrieved power-law slopes around -2.7.

\newpage

\subsubsection{Secondary Tail Evolution} \label{subsubsec:secondary}

As mentioned in Sections \ref{subsec:tail} and \ref{subsubsec:primary}, we retrieved measurements of the secondary tail from both combined and independently resolved brightness profiles. The combined profiles, previously referred to as skewed in Section \ref{subsec:tail}, were observed from 03 to 07 October, when the angular separation of the primary and secondary tails were not large enough to independently resolve them through the atmospheric seeing conditions of each night. We were able to fully resolve the two tails on 14 October as two independent profiles, whose angular separation grew as radial distance from the system increased. Irrespective of the nature of the profile, we find that the secondary tail increased in separation from the primary from 03 to 07 October and that the position angle is on average 2-3$^{\circ}$ higher than the primary tail on the same date of detection. However, by 14 October, this growth reversed, and the secondary tail was slightly closer to the primary. We cannot comment on this trend further since our observations from 19 October have seeing values 5 $\times$ higher than the previous night, which obscured any differentiation between the two tails. The closest 5-15$\arcsec$ to Didymos exhibit curvature in the same direction as the primary tail, suggesting that both tails are undergoing similar gravitational interactions. 

\begin{figure}[ht!]
        \includegraphics[width=\linewidth]{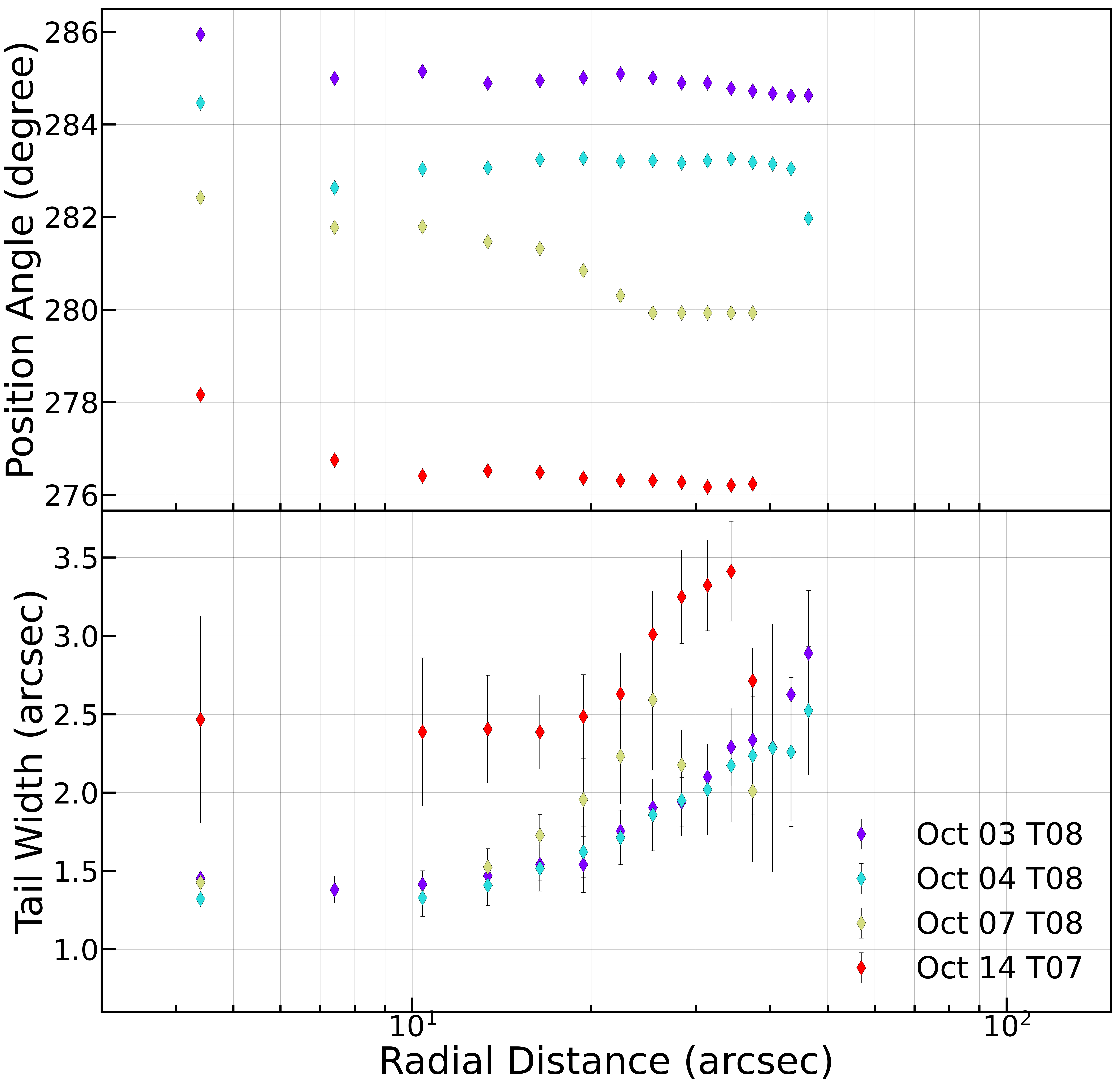}
        \caption{Secondary tail position angles (top panel) and widths (bottom panel), from \textbf{03} to 14 October. Cooler colours represent earlier dates in the campaign, while warmer colours are later dates.\label{fig:t2_meas}}
    \end{figure}

Our width measurements from the secondary tail indicate that it was roughly 1.5-2 $\times$ thicker than the primary tail width retrieved on the same date, as seen in between Figures \ref{fig:t1_width} and \ref{fig:t2_meas}, where at 10$\arcsec$ from the system the primary tail is around 1$\arcsec$ wide and the secondary is around 1.8$\arcsec$ wide. We then matched each secondary tail detection to a previous detection of the primary tail, in order to compare the possible similarities between the two structures at similar times after formation. Synchrones (models representing particles released at the same time) from \cite{Li:2023} suggest the secondary tail likely formed between T+5D to T+7D after impact.  We assumed a median time difference in formation of T+6D, from \cite{Li:2023}, and paired the secondary tail from 03, 04, 07, and 14 October to the primary tail from 27 and 28 September and 01 and 7 October, respectively. We found that secondary tail widths were almost a factor of 2 lower than their primary tail counterparts, which may provide clues about a possible formation mechanism.

Discussions within the DART Investigation Team have identified two primary theories for secondary tail formation -- Sunward-antiSunward ejecta blow-back and a secondary impactor. The former theory posits that a module of ejecta was ejected directly along the Sunward-antiSunward axis and after a few days the module experienced a turnaround point where SRP decelerated the grains and blew them back into the secondary tail. The latter suggests that the initial impact injected large boulders into the local vicinity of Dimorphos, as supported by HST observations of the Didymos boulder swarm \citep{Jewitt:2023}, and some re-impacted the surface of either Dimorphos or Didymos between T+5 to T+7 days post-impact which created a secondary flush of ejecta. Photometric evidence supports this theory, since HST and other optical facilities detected a noticeable pause in dimming between 01 and 03 October in multiple apertures \citep{Li:2023,Rozek:2023,Kareta:2023}, which allows us to infer that new ejecta is likely being excavated rather than old ejecta turning around (that should be within the apertures). Additional Monte Carlo models of tail formation by \cite{Moreno:2023} indicate that transfer of linear momentum, via ejecta impacting, to the surface of both Didymos and Dimorphos exhibit a secondary peak around T+5D post-impact, which may suggest that the secondary tail is the product of numerous secondary impacts across both binary bodies. Lastly, our width measurements hint that a secondary impactor might be more likely, since the blown back ejecta would have a longer time to radially disperse and create a much thicker tail, while the secondary impactor would produce lower velocity ejecta that would create a much thinner tail due to less time to disperse and a much lower ejection energy as compared to the primary impact. Robustly stating that our widths support a secondary impact is nontrivial, as 3 out of our 4 dual tail detections do not have the secondary tail fully resolved, and as a result, the majority of our dual fits are mildly anti-correlated. Despite this anti-correlation, we tested a diverse range of input parameter combinations and we found that our models yielded consistent measurements independent of the input combination, which leads us to believe that the ratios between the primary and secondary tail widths can be trusted. Therefore, we support a secondary impactor as the likely cause for the secondary tail formation.

While our measurements support a secondary impactor formation scenario for the dual tail morphology, a recent paper by \cite{Kim:2023} suggests the secondary tail is not the result of a later impact, but rather the effects of rapidly changing viewing geometry. \cite{Kim:2023} implemented Monte Carlo (MC) simulations of HST observations \citep{Li:2023} that considered the plane-of-sky projection of the hollow ejecta cone with respect to the changing viewing angle. They found that the emergence, separation, and eventual dissipation of the secondary tail could be explained by varying projections of the walls of the hollow ejecta cone over time and viewing angle. Their models retrieve grain size distributions, total ejected mass, position angles, and morphology of the dual tails that are consistent with other studies \citep{Li:2023,Graykowski:2023,Opitom:2023,Daly:2023, Graykowski:2023}. Notably, \cite{Kim:2023} report a jump in scattering cross-section at T+8.8D post-impact, which they correlate to the viewing angle crossing 90$^{\circ}$, at which point the cone walls are most visible. This jump is incongruent with the widely documented T+5-7D post-impact photometric brightening \citep{Li:2023,Kareta:2023,Rozek:2023}, and \cite{Kim:2023} postulate this is due to the decreased photometric sensitivity of ground-based observations. However, the T+5-7D photometric brightening has been robustly characterized by multiple ground and space-based observatories \citep{Li:2023,Kareta:2023,Rozek:2023}, so we posit that the probability that it is an artefact of ground-based observations is highly unlikely. Furthermore, we find a visible difference between the brightnesses of the dual tails, however, the MC models do not appear to reproduce this, which possibly highlights another incongruence between their models and our, and other, datasets \citep{Kareta:2023,Rozek:2023}. Considering these discrepancies, further investigation is required to fully understand the appearance of the secondary tail, however, we remain in support of the secondary impactor formation scenario as evidenced by our data and analyses, and other studies. 

    \begin{figure}[ht!]
        \includegraphics[width=\linewidth]{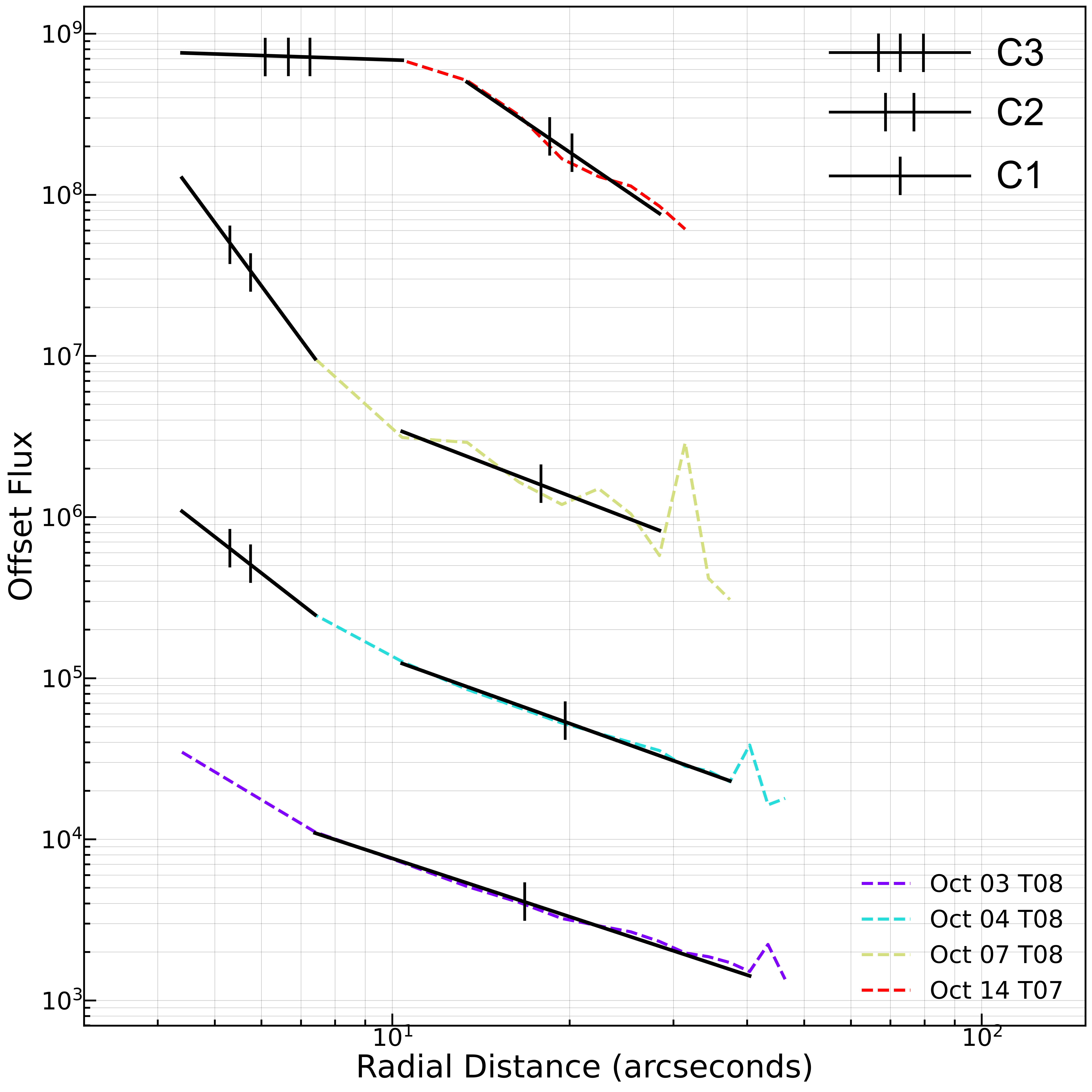}
        \caption{Secondary tail relative flux profiles as measured from the integrated area under the Lorentzian fits. Cooler colours represent earlier dates in the campaign, while warmer colours are later dates. Uncertainties have been removed from the measurements, since we have artificially offset the profiles to compare their relative slopes. Black lines depict the regions where we applied power-law fits. \label{fig:t2_fluxshift}}
    \end{figure}

Following our relative flux comparison for the primary tail, we fit similar features at similar radial distances from Didymos for the secondary tail. As shown in Figure \ref{fig:t2_fluxshift}, a distinct trend can be seen, where the secondary tail brightness morphology mimics the primary tail morphology over a condensed period of time. The secondary tail is seen to have three periods of increasing steepness, extreme steepness, and then a shallow plateau respectively. The fit series are termed C1, C2, C3 range from range from 430-1100 km 03-07 October, 250-850 km 04-07 October, and 250-670km on 14 October and have slopes of -0.8 to -1.2, -2.4 to -3.6, and -0.2, respectively. Intriguingly, we retrieved similar slopes between analogous periods, where C1 is similar to A1, C2 to A2, and C3 to A3. Perhaps this similarity is intrinsic of standard evolutionary phases of both tails, however, the secondary tail is on an accelerated timeline with respect to the primary tail. This accelerated evolution could be the result of the hypothesized lower energy secondary impact, since a low energy impact would eject smaller less massive grains that would be more easily influenced by SRP, which then sped up the overall evolution.

\newpage

\section{Conclusions} \label{sec:conclusion}

We conducted an analysis of post-DART impact observations using the MUSE instrument. Our investigation spanned various spatial regimes and observational timescales, and we investigated the morphological and spectral evolution of the ejecta cone structures and tails up to T+28.3D post-impact. We reported our observational measurements and discussed our findings in the context of other papers and communal discussions, however, we did not carry out detailed modeling as it is outside the scope of this paper. Our findings are as follows:

\begin{itemize}
    \item  We followed the morphological evolution of the ejecta cone from dusty edges to detached wings in the weeks after impact, akin to what is identified by HST \citep{Li:2023} and other studies \citep{Lin:2023, Rozek:2023}. We measured an SRP-induced blow-back velocity of 0.85 m s$^{-1}$ at the base of the northern wing and derived a distribution of maximum grain sizes that could correspond with the velocity. We found that the base of the wing is likely comprised of grains with radii in the range of 0.05-0.2 mm.
    \item We observed 13 unique clumps throughout the ejecta cone in the first days post-impact, in both modes. We calculated a set of velocities between the first and last appearance, first appearance and Didymos, and last appearance and Didymos, which were all consistent and indicative of time-of-impact ejection. Finally, we found that our slowest, innermost clumps in NFM had reflectance slopes 2 $\times$ higher than background slopes, suggesting that they were comprised of slower, larger particles \citep{Fahnestock:2022, Jewitt:1986}. 
    \item We measure the primary tail from just after impact to the end of our observations, and found that the primary tail is curved within the first 5-10$\arcsec$ from the system, which is likely due to binary dynamics. The width of the tail increased with radial distance from the system, but decreased to a minimum around 03 to 04 October. The relative flux intensity slopes are consistent with HST measurements \cite{Li:2023} for the A1 region.
    \item We do not see evidence for the secondary tail until 03 October, where a positive skew was detected in the brightness profiles, which suggests a time of formation after 01 October but before 03 October. These dates bracket the proposed date of formation by \cite{Li:2023}, and the 03 October data are the tightest constraints yet published. We measured the position angles of the secondary tail, and found that it exhibits similar curvature as the primary tail, the secondary tail is consistently more northern than the primary, and that the tails are separating from 03 to 07 October. By 14 October, the dual tail separation is slightly smaller than that of 07 October, however, further characterization of this trend was not possible due to observational conditions on 19 October. Furthermore, the secondary tail is on average thinner than the primary tail at similar times after formation, suggesting that the grains in the tail have lower velocity distributions and therefore might come from a low energy formation mechanism, such as low velocity or low mass secondary impact(s). Finally, we analysed the secondary tail relative flux profiles and identified similar periods with similar slopes as the primary tail, however, these periods evolve over the course of 11D, not 28D. This phenomenon might hint at common evolutionary stages in the formation and evolution of the tail, and that the secondary tail is likely comprised of smaller particles that were more rapidly influenced by SRP.
\end{itemize}

\begin{acknowledgments}
This analysis was based on observations collected at the European Southern Observatory under ESO programmes 110.23XL and 109.2361. The authors similarly wish to acknowledge the dedication and technical support of the support astronomers, Instrument Operators, and Telescope Operators at Paranal Observatory. Furthermore, the authors would like to thank the DART Investigation Team for the opportunity to collaborate on such an outstanding mission, and highlight the exciting work done by all members. This work was supported in part by the DART mission, NASA Contract No. NNN06AA01C to JHU/APL. BM acknowledges funding support by the Royal Astronomical Society for travel costs to conferences and collaborators of this work.
\end{acknowledgments}

\facilities{VLT:Yepun}
\software{IRAF, MPDAF, CCIEF}

\newpage

\appendix

\section{Data Accessibility}  \label{sec:data access}

We affirm that our research data will be made accessible to the scientific community as a data product upon the publication of our findings. This commitment to data accessibility is in line with the principles of transparency and reproducibility, our research team, and the DART collaboration, facilitating the validation of our results and encouraging further scientific inquiry. Our data product will consist of our flux-calibrated datacubes, white-light images, stacked images, enhanced and unenhanced polar and Cartesian images, clump velocities, cone position angles, primary and secondary tail position angles, widths, fluxes, power-law slopes of primary and secondary tail fluxes, and relative reflectance maps. The raw uncalibrated data products can be accessed via the European Southern Observatory's User Access Portal, under programmes 109.2361 and 110.23XL.

\bibliography{sample631}{}
\bibliographystyle{aasjournal}

\end{document}